\documentclass[%
 reprint,
 amsmath,amssymb,
 aps,
prb,
]{revtex4-2}

\usepackage{graphicx}
\usepackage{dcolumn}
\usepackage{bm}
\usepackage{siunitx}

\begin{document}

\preprint{APS/123-QED}

\title{Optimizing density-functional simulations for two-dimensional metals}
\author{Kameyab Raza Abidi}
 \author{Pekka Koskinen}%
 \email{pekka.j.koskinen@jyu.fi}
\affiliation{%
 NanoScience Center, Department of Physics, University of Jyväskylä, 40014 Jyväskylä, Finland
}%

\date{\today}

\begin{abstract}
Unlike covalent two-dimensional (2D) materials like graphene, 2D metals have non-layered structures due to their non-directional, metallic bonding. While experiments on 2D metals are still scarce and challenging, density-functional theory (DFT) provides an ideal approach to predict their basic properties and assist in their design. However, DFT methods have been rarely benchmarked against metallic bonding at low dimensions. Therefore, to identify optimal DFT attributes for a desired accuracy, we systematically benchmark exchange-correlation functionals from LDA to hybrids and basis sets from plane waves to local basis with different pseudopotentials. With 1D chain, 2D honeycomb, 2D square, 2D hexagonal, and 3D bulk metallic systems, we compare the DFT attributes using bond lengths, cohesive energies, elastic constants, densities of states, and computational costs. Although today most DFT studies on 2D metals use plane waves, our comparisons reveal that local basis with often-used PBE exchange-correlation is well sufficient for most purposes, while plane waves and hybrid functionals bring limited improvement compared to the greatly increased computational cost. These results ease the demands for generating DFT data for better interaction with experiments and for data-driven discoveries of 2D metals incorporating machine learning algorithms.
\end{abstract}

\maketitle


\section{\label{sec:intro}Introduction}

The discovery of graphene nearly two decades ago sparked an entire new research field of two-dimensional (2D) materials \cite{graphene}. The 2D materials pedigree has expanded ever since, thanks to unique properties and visions for novel applications \cite{2D_atomic, TMD_app, application2D, 2D_app}. Most 2D materials are covalently bound and have layered structures easily exfoliable from three-dimensional (3D) bulk matter \cite{TMD, BeyondTMD}. However, in contrast to directional covalent bonding, non-directional metallic bonding prefers large coordination numbers, which renders low-dimensional metal structures energetically unfavourable. Despite this preference for large coordination, in 2014 atomically thin stable iron patches were discovered in graphene pores \cite{zhao_free-standing_2014}. This discovery has been followed by rapid progress in research on 2D metals and alloys, making 2D metals a full member the 2D materials family \cite{review1,review2,review3, review4, CuAu, Au}.  

The wavering stability of 2D metals makes experiments challenging, whereby research relies heavily on computations. A reasonable description of metallic bonding requires electronic structure simulations, which has made the density-functional theory (DFT) \cite{hohenberg_inhomogeneous_1964,kohn_self-consistent_1965} the workhorse method for modeling 2D metals \cite{yang_new_2015, yang_glitter_2015,yang_properties_2016,nevalaita_atlas_2018, nevalaita_beyond_2018, nevalaita_stability_2019,nevalaita_free-standing_2020,ono_dynamical_2020,ren_magnetism_2021,ono_comprehensive_2021,Anam_2021, Kapoor_2021, Gallenene, Sangolkar_2022, sangolkar_density_2022}. Most DFT studies have chosen plane wave (PW) basis sets \cite{PW} and the non-empirical Perdew-Burke-Ernzerhof (PBE) exchange-correlation functional \cite{PBE}. These choices for DFT attributes are plausible in the context of delocalized electrons in periodic systems that are still lacking experimental data. However, DFT attributes have not been systematically benchmarked for metallic bonding at low dimensions. It is not certain whether these standard choices are efficient and accurate enough or they if simply waste computational resources.

The DFT attributes consist of few central choices. The first choice is the flavor of exchange-correlation (xc) functional, the level of which is of central importance for consistent results. A functional performing well in some systems may perform poorly in others. Here we make use of several xc-functionals to obtain a systematic picture of their performance in low-dimensional metallic bonding \cite{LibXC}. The second choice is the type of basis function. Plane waves are suitable for periodic systems, whose electrons fill out the entire simulation cell. Unfortunately, the non-periodic directions of low-dimensional systems require large vacuum regions that make PW simulations inefficient compared to modeling bulk. Thus, an additional choice in PW simulations is an optimum size of the vacuum. In this respect, PW and grid-based DFT share the same challenges \cite{grid1, GPAW}. Another alternative for basis is linear combination of atomic orbitals (LCAO), and controlling its size provides a powerful handle to trade between accuracy and efficiency \cite{LCAO}. 

The choice of basis type has implications beyond mere accuracy.
For example, PW is not suitable for studying electron transport using nonequilibrium Green's function method in nanoscaled devices \cite{NEGF}. In addition, with the coming of data science and machine learning in materials science, lots of consistent DFT data is required for machine learning -enabled 2D metals studies \cite{ML, Ml_1, ML-1, ML-2, ML-3}. This efficiency demand calls for a critical examination of the necessity of PW method to model metallic bonding in low dimensions.

Third choice for periodic systems is the number of k-points along periodic directions for the desired accuracy. Fourth choice is the level of Fermi-broadening of electronic states, which is partly a physical choice but mostly a necessity for rapid convergence of the self-consistent iteration of the electron density. In practice, there are a plethora of other choices to make for numerical stability and speedup, but they are often chosen as default values that have been previously fine-tuned for each DFT code.
\begin{figure}[t!]
\centering
\includegraphics[width= 8.6 cm]{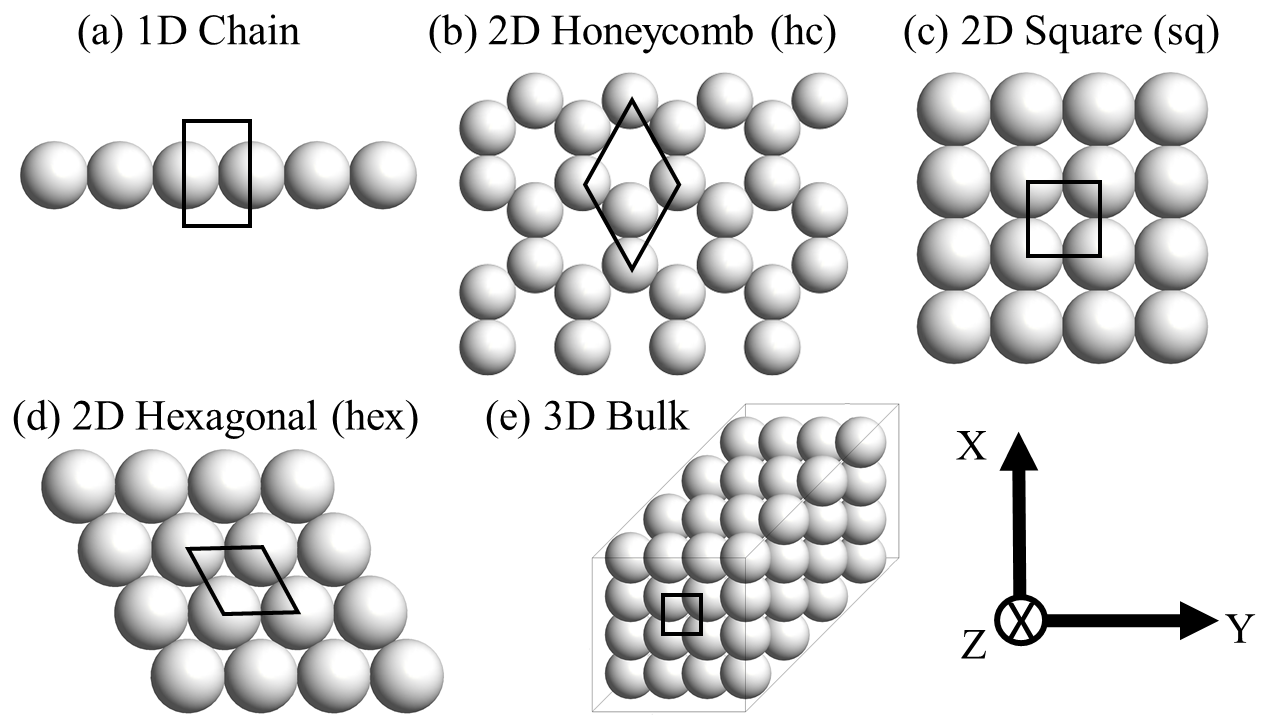}
\caption{Schematics of the systems with different dimensionalities and coordination numbers $C$: 1D chain ($C=2$), 2D honeycomb ($C=3$), 2D square ($C=4$), 2D hexagonal ($C=6$), and 3D bulk ($C=12$). The quadrilaterals show the simulation cells.}
\label{fig:systems}
\end{figure}
In this article, we consider the above-mentioned choices of DFT attributes regarding xc-functionals, basis sets, vacuum, k-point sampling, and Fermi-broadening, and juxtapose their performance against various properties of selected low-dimensional metal systems. The selected systems include a one-dimensional chain (coordination number $C=2$), three two-dimensional lattices ($C=3$, $4$, and $6$), and a 3D bulk ($C=12$) (Figure \ref{fig:systems}). These systems enable comparative analysis of the performance of DFT attributes in various dimensions. Being low-dimensional systems, these structures are prone to various symmetry-breaking deformations, such as out-of-plane buckling in 2D or Peierls distortions in 1D \cite{ono_comprehensive_2021, Peierls}. However, in order to enable unambiguous comparison of the effect of dimensionality and coordination and avoid making unfounded conclusions based on incomplete set of deformations, we retain our focus on these ideal, non-deformed systems. We also compare the performance and speed of DFT to the density-functional tight-binding (DFTB) method, which is the next-in-line approximation to DFT \cite{DFTB}. One of our main conclusions is that, for general purposes, DFT-LCAO can be chosen over the default DFT-PW without compromising accuracy, a choice which enables simulating transport and helps generating DFT data more effortlessly. Our treatise will advance DFT modeling of 2D metals and help boosting the interaction with experiments.

\section{\label{sec:2}Computational Methods}
The basic idea DFT is to use the variational principle to generate exact ground state energy and density for the systems of interest \cite{hohenberg_inhomogeneous_1964}. The ground state energy $E$ is a functional of the electron density (\textit{n}),
\begin{equation}
\label{eq:1}
E[n] = T[n]  + E_{ext}[n]+ E_{H}[n] + E_{xc}[n]\, ,
\end{equation}
where $T[n]$ is the Kohn-Sham kinetic energy for the fictitious non-interacting electron system, $E_{ext}[n]$ is the external potential energy, $E_{H}[n]$ is the Hartree energy, and $E_{xc}[n]$ is the exchange-correlation energy. The xc term attempts to capture the complex features of many-body quantum mechanics, and a variety of approximate xc functionals have been developed for different purposes \cite{LibXC}. As a result, the quality of xc functional mostly determines the quality of the results. Here, using the QuantumATK (S-2021.06) DFT implementation \cite{QuantumATK}, we explore the set of eight xc functionals ranging from local density approximation to hybrid functionals (Table \ref{tab:xc}).

We used two types of basis sets, plane waves and LCAOs. The wave-function energy cutoff for plane waves was 800 \si{\eV}. Cutoff needed no separate analysis for low-dimensional metals, because it depends only on element and pseudopotential \cite{pse}. For LCAOs, we used three variants: LCAO-M(edium), LCAO-H(igh), and LCAO-U(ltra). These variants derive from the numerical basis sets of the FHI-aims package \cite{FHI}, but are further optimized for computational speed of the LCAO calculator. For example, for Ag the radial functions for Medium basis are $3s/2p/1d$ (14), for High $4s/3p/5d/1f$ (35), and for Ultra $4s/3p/5d/2f/1g$ (51), with brackets displaying the total number of orbitals per atom \cite{LCAO, FHI}. Local basis sets were used in conjunction with  norm-conserving PseudoDojo pseudopotentials \cite{PSeudoDojo}.

Further, the total energy convergence criteria for self-consistent electron density was $\leq{10^{-7}}$\si{\eV}. System geometries were optimized to forces below 1 m\si{\eV\angstrom^{-1}} and stresses below 0.3 m\si{\eV\angstrom^{-3}} using the LBFGS \cite{LBFGS} algorithm. The k-points were sampled by the Monkhorst-Pack method \cite{Monkhorst_Pack}. All calculations were spin-polarized and the initial guess for lattice parameters were adopted from the Atlas of 2D metals \cite{nevalaita_atlas_2018}.

To complement the results with various DFT attributes with wider context, we analyzed the systems with Ag also with DFTB method at the level of self-consistent charge \cite{DFTB,DFTBforbeginner}. The Ag parametrizations were taken from earlier studies \cite{DFTB_2, DFTB_1}.

\begin{table}[t!]
\begin{ruledtabular}
\centering
\caption{
Exchange-correlation functionals used in this work.}
\begin{tabular}{ll}
\multicolumn{1}{l}{\bf Functional and its family}  & \multicolumn{1}{l}{\bf Refs.}\\
\hline
\multicolumn{1}{l}{Local Density Approximation (LDA)} &  \multicolumn{1}{l}{\cite{hohenberg_inhomogeneous_1964, LDAPW}}\\
\multicolumn{1}{l}{ Generalized Gradient Approximation (GGA)} & \multicolumn{1}{l}{\cite{GGA}}\\
\multicolumn{1}{l}{\hspace{0.3cm}RPBE} & \multicolumn{1}{l}{\cite{RPBE}}\\
\multicolumn{1}{l}{\hspace{0.3cm}PW91} & \multicolumn{1}{l}{\cite{pw91_1,PW91}}\\
\multicolumn{1}{l}{\hspace{0.3cm}PBE} & \multicolumn{1}{l}{\cite{PBE}}\\
\multicolumn{1}{l}{Hybrid Functionals} & \multicolumn{1}{l}{\cite{Hybrid}}\\
\multicolumn{1}{l}{\hspace{0.3cm}B3LYP} & \multicolumn{1}{l}{\cite{B3LYP}} \\
\multicolumn{1}{l}{\hspace{0.3cm}PBE0} & \multicolumn{1}{l}{\cite{PBE0}} \\
\multicolumn{1}{l}{\hspace{0.3cm}HSE03 (screening $\omega$ = 0.15 $Bohr^{-1}$)} & \multicolumn{1}{l}{\cite{HSE03_06}}\\
\multicolumn{1}{l}{\hspace{0.3cm}HSE06 (screening $\omega$ = 0.11 $Bohr^{-1}$)} & \multicolumn{1}{l}{\cite{HSE}}
\end{tabular}
\label{tab:xc}
\end{ruledtabular}
\end{table}%

\section{\label{sec:3}Results and Discussion}
\subsection{\label{subsec:3a}Convergence Analysis}
We made various systematic convergence analyses for the group of coinage metals Cu, Ag, and Au \cite{yang_new_2015, yang_glitter_2015, yang_properties_2016}. Computational and experimental studies have shown that the free-standing monolayer patches of these metals are stabilized by graphene pores \cite{ono_dynamical_2020, sangolkar_density_2022, nevalaita_stability_2019, CuAu}. The analyses were done using PBE xc-functional \cite{PBE}, projector augmented waves (PAW) for core electrons \cite{PAW}, and plane waves for valence electrons.

\paragraph{k-point convergence:}
The k-point convergence was studied using the 2D systems with a converged vacuum of 15~\si{\angstrom} in the non-periodic direction (as confirmed below). The total energy is practically converged at $30\times30\times1$ k-point sampling, and we define the energy tolerance using this value, 
\begin{equation}
\label{eq:2}
    \Delta{E} = E_{N_{k}\times N_{k}\times1} - E_{30\times30\times1} \, .
\end{equation}
Apart from rapid convergence at very few k-points, the convergence is exponential. Chosen relative energy tolerance can therefore be approximated by 
\begin{equation}
    \log\delta = A_{1} + B_{1}L \, ,
\label{eq:k-conv}
\end{equation}
where \(\delta = \mid\Delta{E}\mid/E\textsubscript{3D} \) is an (approximate) relative energy tolerance, the ratio between energy tolerance to the 3D cohesive energy \( E\textsubscript {3D} \) \cite{kittel}. The length $L=a_{c} N_k$, the product of simulation box length and the number of k-points in corresponding direction, is the maximum period of the Bloch wave function. Using $L$ as the convergence parameter helps identifying the required k-point sampling for variable simulation cell sizes in later research.

\begin{figure}[t!]
\centering
\includegraphics[width=8.6cm]{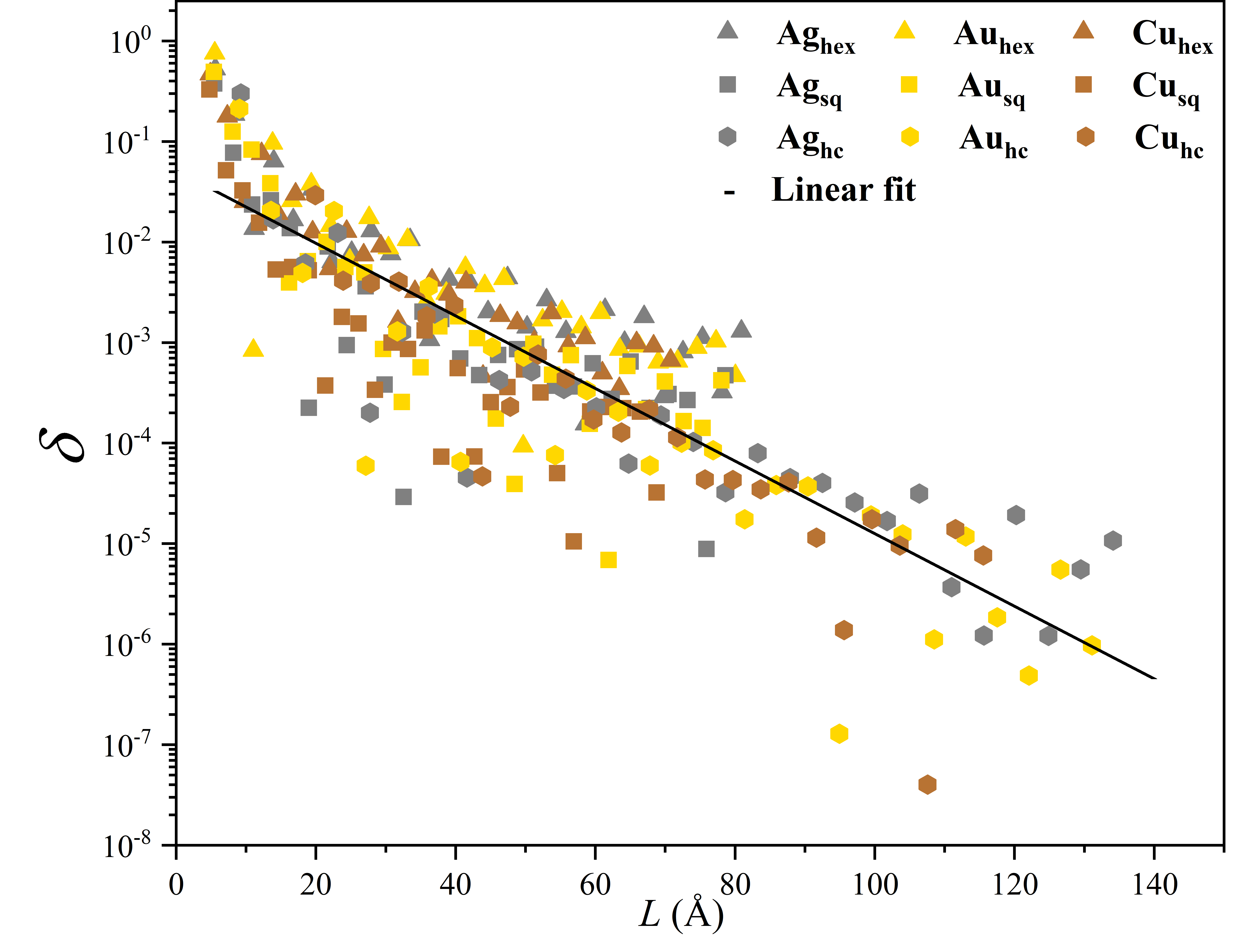}
\caption{The k-point convergence of total energy for 2D systems made of coinage metals. $\delta$ is the relative energy tolerance and $L$ is the maximum period of the Bloch function [\emph{cf}. Eq.(\ref{eq:4})]. The linear fit refers to Eq.~(\ref{eq:k-conv}).}
\label{fig:k-point}
\end{figure}

The k-point convergence is not monotonic; more k-points does not necessarily mean better accuracy (Figure \ref{fig:k-point}). However, for different system symmetries and cell shapes and sizes, the ansatz (\ref{eq:k-conv}) works satisfactorily. Linear regression analysis to the data gives the parameters \( A_{1} = -1.29\) and \( B_{1} = -0.036 \; \;\si{ \angstrom^{-1}}\) (Figure \ref{fig:k-point}). Inverting Eq.~(\ref{eq:k-conv}), we can obtain an optimal number of k-points for given simulation cell size $a_c$ and desired accuracy $\delta$ as
\begin{equation}
\label{eq:4}
    N_{k}(\delta) = \text{ceil}\left(\frac{L(\delta)}{a_c}\right), 
\end{equation}
where ceil(\begin{math} x \end{math}) = \begin{math}\lceil x \rceil\end{math} maps $x$ to the least integer greater than or equal to \begin{math} x \end{math}. For instance, with relative accuracy \(\delta =10^{-3}\) one obtains the \(N_{k}\) = \(\lceil 47 \text{ \AA}/a_c \rceil\), suggesting $\Gamma$-point calculations for 4.7-nm-sized simulation cells. In subsequent analyses, we use \begin{math} N_{k} = 13 \end{math}, suggesting \(\sim \delta = 10^{-2.5 \ldots -3}\) relative tolerance.

\paragraph{Vacuum convergence:}
Using plane waves requires periodicity in all directions, regardless of system dimensions. 
Low-dimensional systems need therefore a large vacuum region in the non-periodic direction to avoid spurious interactions with periodic images of the system. Larger vacuum means more volume and computational cost, implying a need to minimize the vacuum without affecting the energy. For a complete picture, we investigate vacuum convergence not only in 2D systems and but also in 1D chains and free atoms. 

\begin{figure}[t!]
\centering
\includegraphics[width= 8.6 cm]{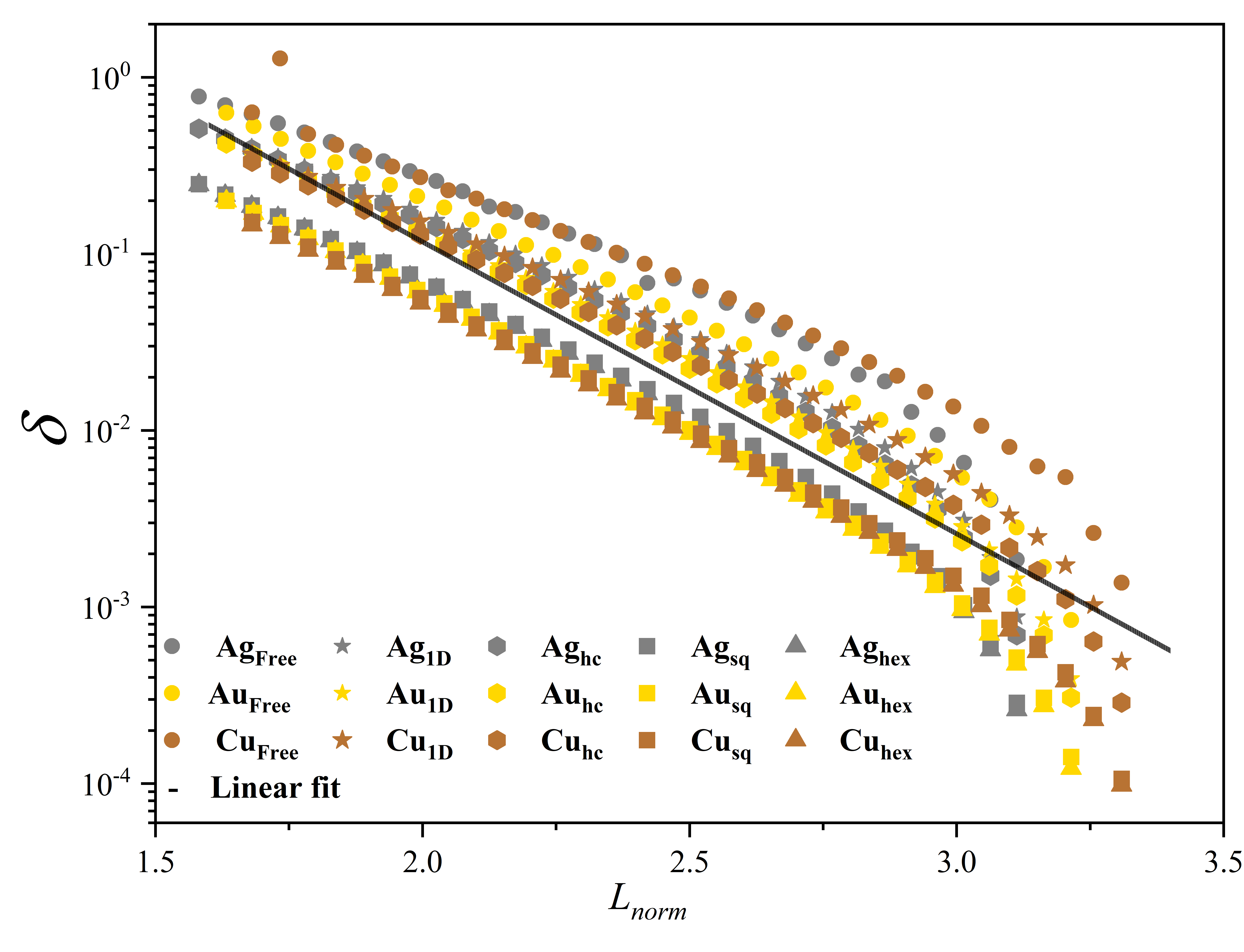}
\caption{Vacuum convergence of the total energy for 1D and 2D systems made of coinage metals. $\delta$ is the relative energy tolerance and $L_\text{norm}$ is vacuum normalized in terms of van der Waals radii. Free atom vacuum convergences are added for comparison.}
\label{fig:vacc-conv}
\end{figure}

We normalize atoms' dimensions by their van der Waals radii \(R_{vdW}\) and consider the normalized vacuum \(L_{norm} = L_{vac}/R_{vdW}\), where \(L_{vac}\) is the vacuum along the non-periodic direction (\emph{i.e.}, the separation between periodic images.) The total energy is practically converged at $8$-\AA\ vacuum, and we define the energy tolerance as \(\Delta E = E(L_{vac}) - E({\si{8\; \angstrom}})\) and relative energy tolerance again as $\delta=\Delta E/E\textsubscript{3D}$. The tolerance converges roughly exponentially, $\log \delta = A_2 + B_2 L_\text{norm}$ (Figure \ref{fig:vacc-conv}). Consequently, the vacuum for a desired relative energy accuracy for a given element can be estimated from 
\begin{equation}
\label{eq:5}
    L_{vac}(\delta) = R_{vdW}\frac{(\log\delta -A_{2})}{B_{2}}\,,
\end{equation}
where the parameters \(A_{2} = 2.38\) and \(B_{2} = -1.65\) were obtained by linear regression. For instance, the relative tolerance \(\delta =10^{-3}\) requires \(L_{vac} = 3.3\times R_{vdW}\). In subsequent analysis, if not said otherwise, we will use \(L_{vac} = \si{10\; \angstrom}\), which for Ag means \(\delta =10^{-4.2}\), in rough alignment with k-point convergence.

Still, such a single estimate is indicative at best. The vacuum convergence follows roughly the coordination number, free atom converging the slowest, hexagonal system the fastest (Figure \ref{fig:vacc-conv}). This suggests that for a given element the vacuum should be set by the lowest-coordinated atom---or by the free atom to be on the safe side. After all, a modest 16~\%\ increase in vacuum ($L_\text{norm}=2.5\rightarrow 3.0$) may increase the relative accuracy by an order of magnitude. Thus, a single fit as above is not the best guideline and the vacuum convergence is best considered by case basis, especially in the presence of possible charge transfer.

\subsection{\label{subsec:3a} Effect of Fermi broadening}
In principle the Fermi-broadening is a physical parameter intimately linked to the electronic temperature $T$; in practice it is frequently used as a technical parameter to accelerate the self-consistency convergence. The technical attitude towards broadening is evident in available methods other than the Fermi-function. Computational literature shows a plethora of different values for Fermi-broadening, but its effect is rarely discussed in detail. For insulators and semiconductors the broadening is inconsequential, but for metals it matters. In this section, we want to investigate its effect on the energetics systematically, for sheer completeness and future reference. 

\begin{figure}[t!]
\centering
\includegraphics[width=8.6 cm]{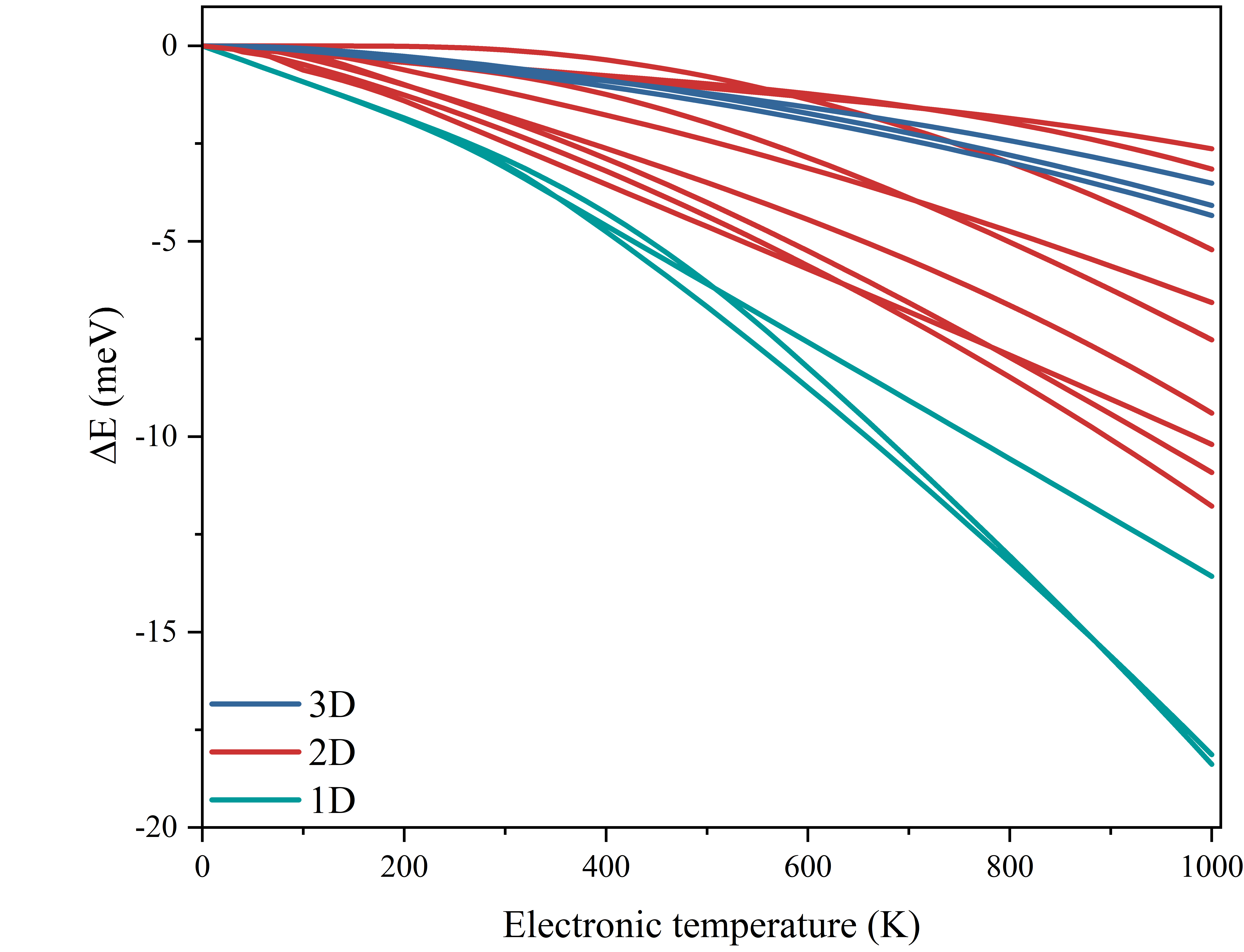}
\caption{The effect of electronic temperature on the cohesion energy of coinage metals in different dimensions.}
\label{fig:broad}
\end{figure}

Ideally, broadening should be chosen to enable rapid convergence without conflicting too much with other convergence parameters. We investigated the effect of broadening by increasing the electronic temperature $T$ from $10^{-5}$~K to $1000$~K and looked at the energy difference 
\begin{equation}
\Delta E(T) = E(T) - E(10^{-5}\text{ K}).
\end{equation} 
The temperature $10^{-5}$~K was the smallest that enabled robust convergence for all systems. Vacuum was $15$~\AA\ for all systems. As a result, 1D systems were most sensitive to the broadening, 3D bulk systems were least sensitive (Figure \ref{fig:broad}). This result is plausible, because the density of states is the smallest for 1D systems. In 2D and 3D systems there are more k-points, density of states at Fermi-level is greater, and state occupations average over a larger set of states, consequently diminishing the influence of broadening. The 2D systems show energy variation around $\sim 10$~meV upon increasing temperature to 1000~K, corresponding to $86$~meV energy broadening (Figure \ref{fig:broad}). For the remainder of the calculations in this article, we used the electronic temperature of $580$~K ($\widehat{=} 0.05$~eV).

\subsection{{\label{subsec:3b}}Performance of exchange-correlation functionals}

We investigated the performance of xc functionals by first fixing certain attributes. To eliminate uncertainties from an insufficient description of valence electrons, we used the most complete PW basis set and the PAW potential to describe the core electrons. We used the converged number of k-points and size of vacuum from previous analysis, as well as the recently adopted $0.05$~eV broadening. With these choices, we may concentrate on the performance of xc-functionals without worrying too much about artifacts from other sources. 

We also investigate xc functionals by using only Ag systems. By belonging to the same group, the coinage metals follow similar trends and it is reasonable to expect other metals to follow the trends of Ag. Still, we do not claim Ag displays completely universal trends, for there are elements that have complex many-body effects even beyond the capabilities of DFT.

In the following, we compare the xc-functional performance against bond lengths, cohesive energies, and elastic moduli of all 1D, 2D, and 3D systems. The electronic structure is compared in terms of later-introduced characteristic figures related to the density of states at the Fermi-level.

\vspace{0.2cm}
\paragraph{Cohesive Energies:}
The cohesive energy was defined as
\begin{equation}
\label{eq:7}
E_{coh} = E_{free} - E/N\, ,
\end{equation}
where \(E\) is the energy of the system with $N$ atoms and \( E_{free} \) is the energy of free atom calculated by placing it inside a 15-\si{\angstrom} cube.

\begin{figure}[t!]
\centering
\includegraphics[width=8.6 cm]{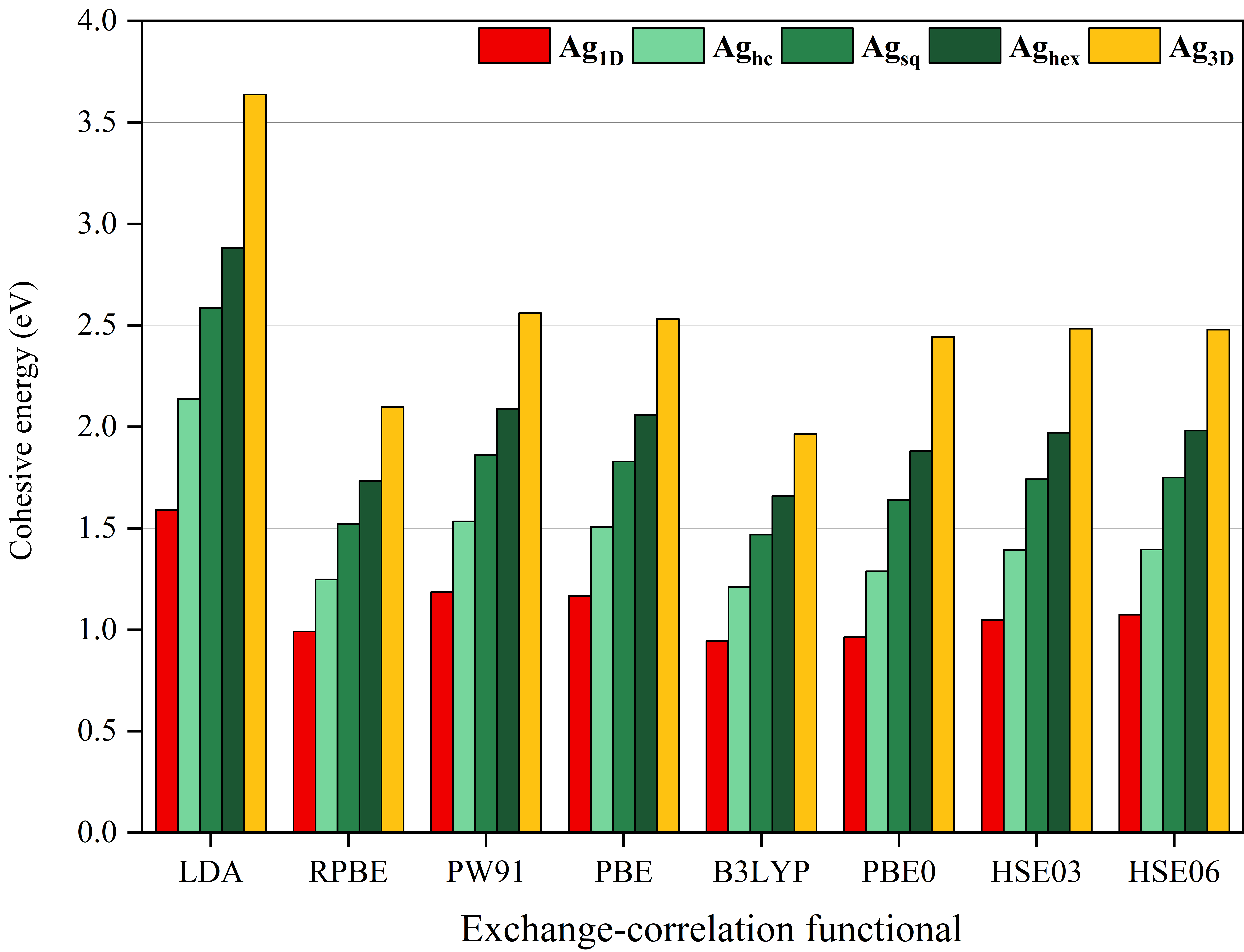}
\caption{The cohesive energies of optimized 1D, 2D (hc, sq, and hex), and 3D systems of Ag with different xc-functionals.}
\label{fig:coh_xc}
\end{figure}

All functionals display similar trends, cohesive energy increasing monotonically from 1D to 3D bulk (Figure \ref{fig:coh_xc}). Yet the quantitative differences are visible. LDA displays its well-known tendency to overestimate cohesive energies. The 3D bulk cohesion shoots over the experimental value by 23~\%\ \cite{kittel}. GGA functionals work significantly better, where PW91 and PBE are now off by approximately $\approx 13-14$~\%. In contrast, RPBE shows considerable underbinding and even less accurate cohesion than LDA. Among hybrid functionals, the performance of screened exchange HSE03 and HSE06 is better than PBE0, which still suffers from the spurious Coulomb interaction. B3LYP describes cohesion poorly and is outperformed by practically all other functionals, and should be avoided while modeling 2D metals---a conclusion not surprising in the light of previous observations \cite{B3LYPfails}. In addition, convergence of free atom with B3LYP was difficult and required loosening the convergence criterion to $\leq{10^{-6}}$~\si{\eV} (loosening had an insignificant effect on the cohesion of Figure~\ref{fig:coh_xc}). As a rule, GGA and hybrid functionals outperform LDA, but a hybrid functionals do not necessarily outperform GGA. PW91 and PBE appear as still as fair choices for robust energetics for general purposes.

\vspace{0.2cm}
\paragraph{Dimensionality-dependence of energetics:}

In 2D metal modeling, the coordination of single metal atoms can range from $C\sim 1$ to $C\sim 6$ and occasionally beyond. The computational method should therefore capture correctly the \emph{relative energetics} of atoms at different coordination numbers. In other words, the cohesion should increase with the coordination number with an appropriate dependence. Our ansatz for the $C$-dependence for the cohesion $E_\text{coh}$ is 
\begin{equation}
    E_\text{coh}(C) = E_\text{coh}^{3\text{D}} \times (C/12)^ \gamma\,,
    \label{eq:gamma}
\end{equation}
where $E_\text{coh}^{3\text{D}}$ is the 3D bulk cohesion and $\gamma$ is an exponent that quantifies the coordination- or dimensionality-dependence of the cohesion energy. The ansatz has the correct asymptotic limits [$E_\text{coh}(0)=0$ and $E_\text{coh}(12)=E_\text{coh}^{3\text{D}}$] and suffices for our purposes in this article. (We tested also more refined ansatzes, but the conclusions remained the same.) The exponent \(\gamma\) was obtained by fitting the Eq.~(\ref{eq:gamma}) for energies from each functional. 

\begin{figure}[t!]
\centering
\includegraphics[width=8.6cm]{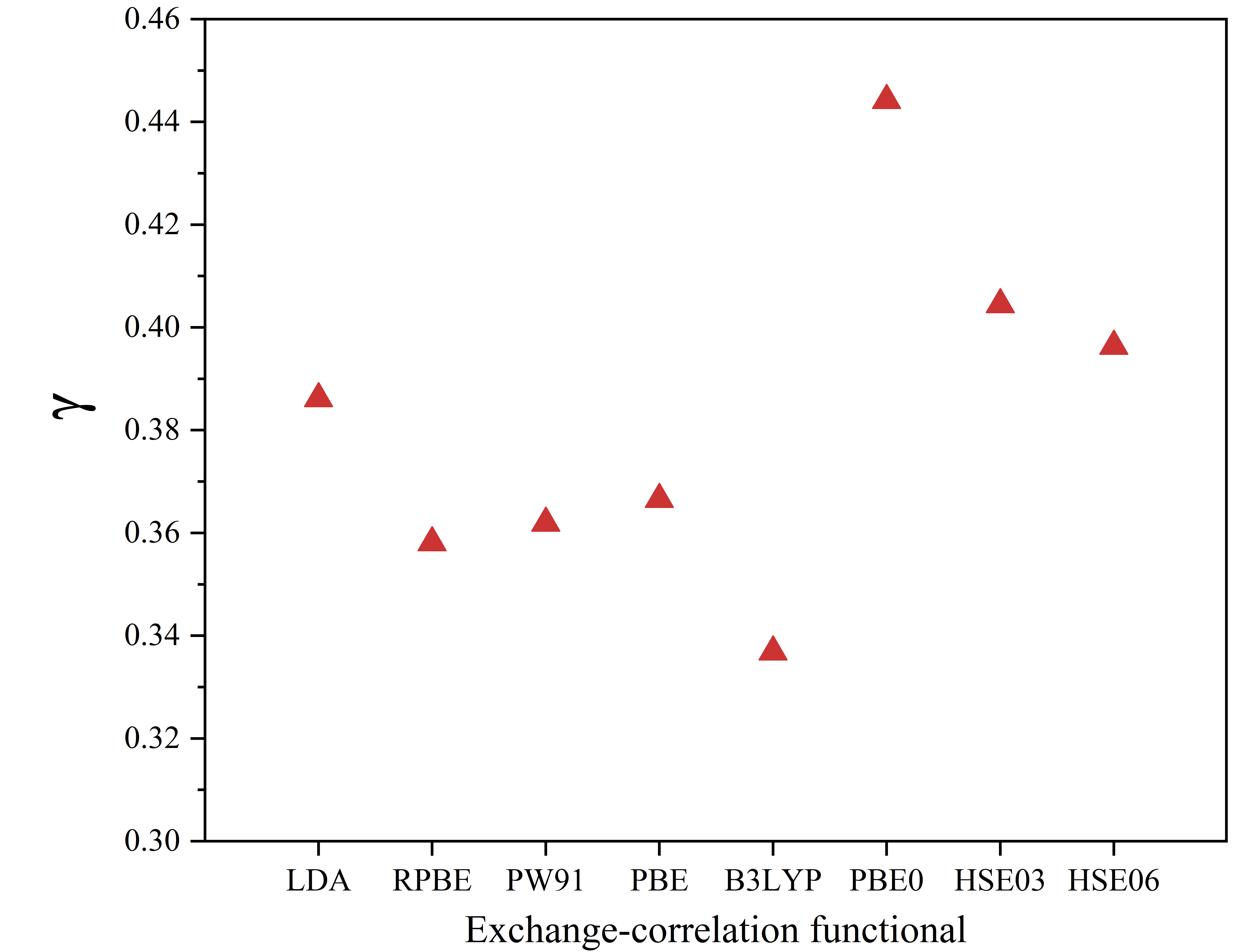}
\caption{Trends of low-dimensional energetics with different xc-functionals. The fitted scaling exponent $\gamma$ is plotted for different xc-functionals; smaller $\gamma$ means that energy depends less linearly on the coordination number [see Eq.(\ref{eq:7})].}
\label{fig:gamm_xc}
\end{figure}

As the result, LDA and all GGA and HSE functionals show roughly the same $\gamma$, the same dimensionality-dependence in energetics (Figure \ref{fig:gamm_xc}). Especially the dependencies in different GGAs are nearly identical. Only the dependencies in B3LYP and PBE0 are clear outliers, PBE0 showing more linear dependence on $C$ ($\gamma$ closer to one) and B3LYP showing more non-linear dependence on $C$ ($\gamma$ further away from one). Interestingly, although LDA badly overestimates the absolute cohesion energies, the dimensionality-dependence lies somewhere in between GGAs and HSE functionals. In conclusion, GGA-PBE appears to capture the dimensionality-dependence of energetics comparably well and be still a serious competitor to the far more costly HSE functionals.

\vspace{0.2cm}

\begin{figure}[b!]
\centering
\includegraphics[width=8.6 cm]{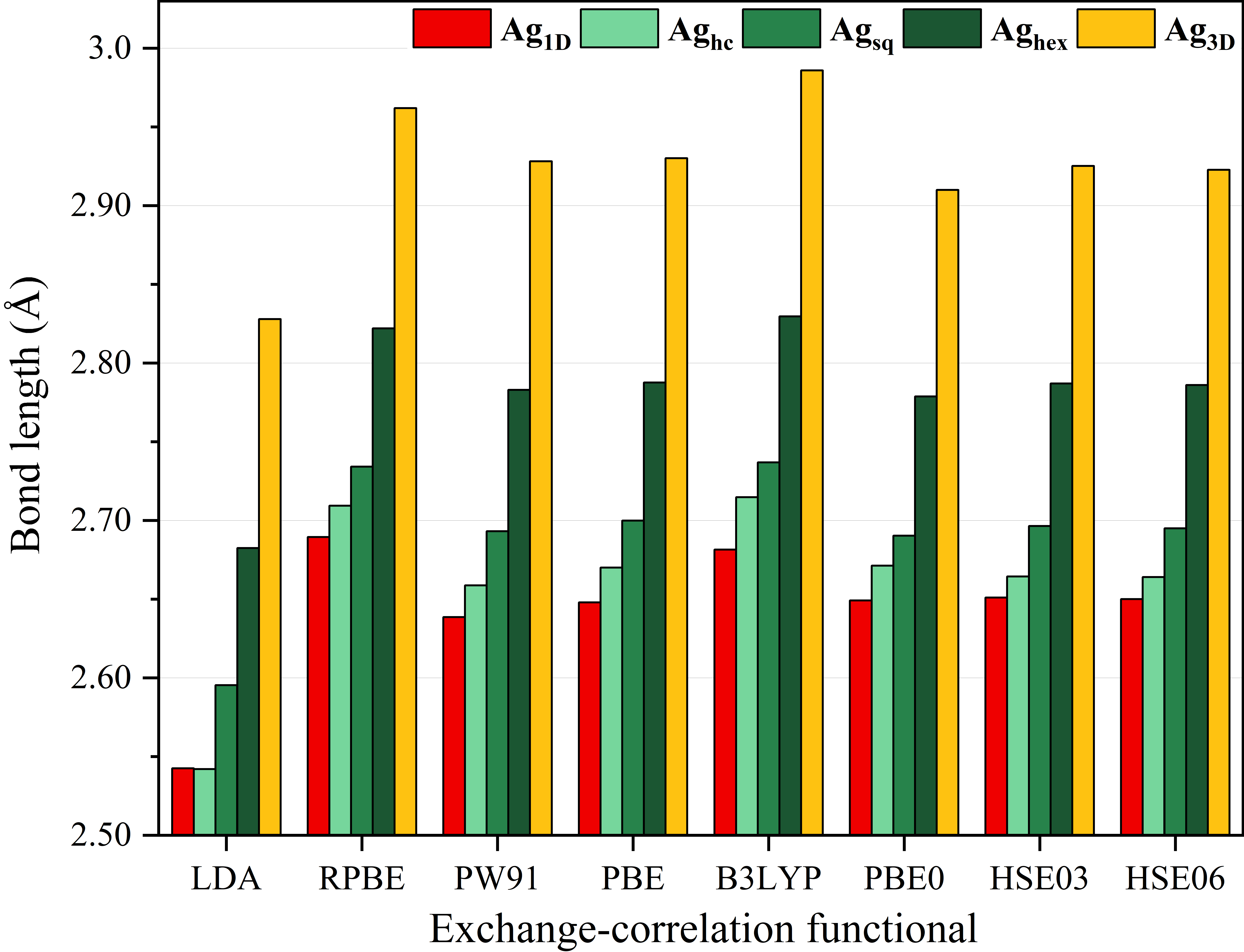}
\caption{Optimized bond lengths of 1D, 2D (hc, sq, and hex), and 3D systems of Ag with different xc-functionals}
\label{fig:bond_xc}
\end{figure}

\paragraph{Bond Lengths:}
The bond lengths were obtained directly from the optimized lattice constants (Figure \ref{fig:bond_xc}). In accordance with overbinding, LDA functional shows small bond lengths. In 3D, the functionals PW91, PBE, PBE0, HSE03, and HSE06 are underbinding and show $1-2$~\%\ too large bond lengths. PBE0 shows shortest bonds among hybrid functionals, and B3LYP shows longest bonds among all functionals. Nearly all functionals show monotonic increase of bond length with coordination number. Only LDA functional is an exception: it has a slightly smaller bond length for 2D hexagonal lattice than for 1D chain.

\vspace{0.2cm}
\paragraph{Elastic constants (theory recap):}
Due to colorful practices in the notations of low-dimensional elasticity, and to avoid any confusion, we wish to define explicitly the elastic constants presented in this article. 


Within the linear elastic regime the stresses \{$\sigma_i$\} and strains \{$\varepsilon_i$\} ($i=1\ldots 6$) satisfy the generalized Hooke's law
\begin{equation}
    \sigma_{i} =  \sum_{j=1}^{6} C_{ij} \varepsilon_{j}\, ,
\end{equation}
where \(C_{ij}\) are elastic constants and expressed as a \(6\times6\) matrix and  \(\varepsilon_{1}= \varepsilon_{xx}, \varepsilon_{2}=\varepsilon_{yy}, \varepsilon_{3}= \varepsilon_{zz},  \varepsilon_{4}= 2\varepsilon_{yz}, \varepsilon_{5}= 2\varepsilon_{xz}, \varepsilon_{6} = 2\varepsilon_{xy}\), when following the Voigt notation. We adapted the formalism of Refs. \cite{Elastic_1, Elastic_2, Elastic_3, Elastic_4, Elastic_Cubic} to evaluate the elastic constants for 1D, 2D and 3D systems.

In 3D, the strain tensor is 
\begin{equation}
\label{eq:8}
\epsilon^{3\text{D}} = \begin{pmatrix} \varepsilon_{1} & \varepsilon_{6}/{2} & \varepsilon_{5}/2\\
\varepsilon_{6}/2 & \varepsilon_{2} & \varepsilon_{4}/2\\
\varepsilon_{5}/2 & \varepsilon_{4}/2 & \varepsilon_{3}
\end{pmatrix}\, .
\end{equation}
The elastic constants are obtained by applying selected strains \{$\varepsilon_i$\} to the equilibrium simulation cell and by calculating the partial derivatives
\begin{equation}
C_{ij} = \frac{\partial^2 \Delta U}{\partial \varepsilon_{i} \partial \varepsilon_{j}}.    
\end{equation}
Here \(\Delta U(\varepsilon_{i}) = U(\varepsilon_{i}) - U(0)\) is the elastic energy density \emph{per unit volume}, where \(U(\varepsilon_{i})\) is the energy density at strain $\varepsilon_i$. For a system with cubic symmetry, the energy density is 
\begin{eqnarray}
\Delta U (\varepsilon_{i}) = && \frac{1}{2} \left( C_{11}\varepsilon_{1}^{2}+C_{11}\varepsilon_{2}^{2}+C_{11}\varepsilon_{3}^{2}+C_{12}\varepsilon_{1}\varepsilon_{2}+ C_{12}\varepsilon_{1}\varepsilon_{3} \right.
\nonumber\\&& 
 +C_{12}\varepsilon_{2}\varepsilon_{1}+C_{12}\varepsilon_{2}\varepsilon_{3}+C_{12}\varepsilon_{3}\varepsilon_{1}+ C_{12}\varepsilon_{3}\varepsilon_{2}
\nonumber\\
&&
 \left. +C_{44}\varepsilon_{4}^{2}+ C_{44}\varepsilon_{5}^{2}+C_{44}\varepsilon_{6}^{2}\right) \, .
\end{eqnarray}

For 2D systems, the strain tensor is
\begin{equation}
\epsilon^{2\text{D}} = \begin{pmatrix} \varepsilon_{1} & \varepsilon_{6}/{2}\\
\varepsilon_{6}/2 & \varepsilon_{2}\\
\end{pmatrix}  \, . 
\end{equation}
Again, the elastic constants are obtained by applying selected strains \{$\varepsilon_i$\} to the equilibrium simulation cell and by calculating the partial derivatives
\begin{equation}
    C_{ij}=\frac{\partial^2 \Delta U}{\partial \varepsilon_{i} \partial \varepsilon_{j}}
\end{equation}
Here \(\Delta U(\varepsilon_{i}) = U(\varepsilon_{i}) - U(0)\) is the energy density \emph{per unit area}, where \(U(\varepsilon_{i})\) is the energy density at strain $\varepsilon_i$. For a system with square symmetry, the energy density is
\begin{eqnarray}
\Delta U (\varepsilon_{i}) = && \frac{1}{2} (C_{11}\varepsilon_{1}^2+C_{22}\varepsilon_{2}^2+2C_{12}\varepsilon_{1}\varepsilon_{2}+ 2C_{16}\varepsilon_{1}\varepsilon_{6}
\nonumber\\&& 
+ 2C_{26}\varepsilon_{2}\varepsilon_{6}+C_{66}\varepsilon_{6}^2)\, 
\end{eqnarray}
and all three elastic constants \(C_{11}\), \(C_{12}\) and \(C_{66}\) are independent. However, for a hexagonal system, only constants \(C_{11}\) and \(C_{12}\) are independent and \(C_{66} = (C_{11}-C_{12})/2\). 

Finally, for 1D systems, the strain-tensor matrix is simply \(\epsilon^{1\text{D}}=(\varepsilon_{1})\). Yet again, the elastic constant is obtained by applying the strain $\varepsilon_1$ to the equilibrium simulation cell and by taking the partial derivative
\begin{equation}
    C_{1}=\frac{\partial^2 \Delta U}{\partial^2 \varepsilon_1}.
\end{equation}
Here \(\Delta U(\varepsilon_{i}) = U(\varepsilon_{i}) - U(0)\) is the energy density \emph{per unit length}, where \(U(\varepsilon_{i})\) is the energy density at strain $\varepsilon_i$. In other words, 
\begin{equation}
\Delta U (\varepsilon_{1}) =  \frac{1}{2} C_{11}\varepsilon_{1}^2\, .
\end{equation}

Table~\ref{tab:elastic_moduli} summarizes the formulae for the elastic constants and their relations. Note that the elastic constants in different dimensions have also different units: they are GPa for 3D, GPa nm for 2D, and GPa nm$^{2}$ for 1D (GPa nm$^{3-D}$ or eV/\AA$^D$ in short, where $D$ is the dimensionality).

\begin{table}[t!]
\caption{\label{tab:1}
Formulae for Bulk Modulus (K), Shear-modulus (G), Young's modulus (Y), and Poisson's ratio ($\mu$) for the systems in Fig.~\ref{fig:systems}.
}
\begin{ruledtabular}
\begin{tabular}{lccccc}
\label{tab:elastic_moduli}
\textrm{System}&
\textrm{K}&
\textrm{G}&
\textrm{Y}&
\textrm{$\mu$}\\
\colrule\\
1D & \(C_{11}\) & - & \(K\) & - \\
\\
2D\textsubscript{hex/hc} & \( \frac{C_{11}+C_{12}}{2}\) & \( \frac{C_{11}-C_{12}}{2}\)& \(\frac{4KG}{K+G}\)& \(\frac{K-G}{K+G}\)\\
\\
 2D\textsubscript{sq}& \( \frac{C_{11}+C_{12}}{2}\) & \( C_{66}\)& \( \frac{C_{11}^{2}-C_{12}^{2}}{C_{11}}\) &\(\frac{C_{11}}{C{12}}\) \\
\\
3D &\( \frac{C_{11}+2C_{12}}{3}\) & \( \frac{3C_{44}+C_{11}-C_{12}}{5}\) & \(\frac{9KG}{3K+G}\)&\(\frac{3K-2G}{2(3K+G)}\) \\

\end{tabular}
\end{ruledtabular}
\end{table}
\begin{figure}
\centering
\includegraphics[width=8.6cm, height=22.1cm]{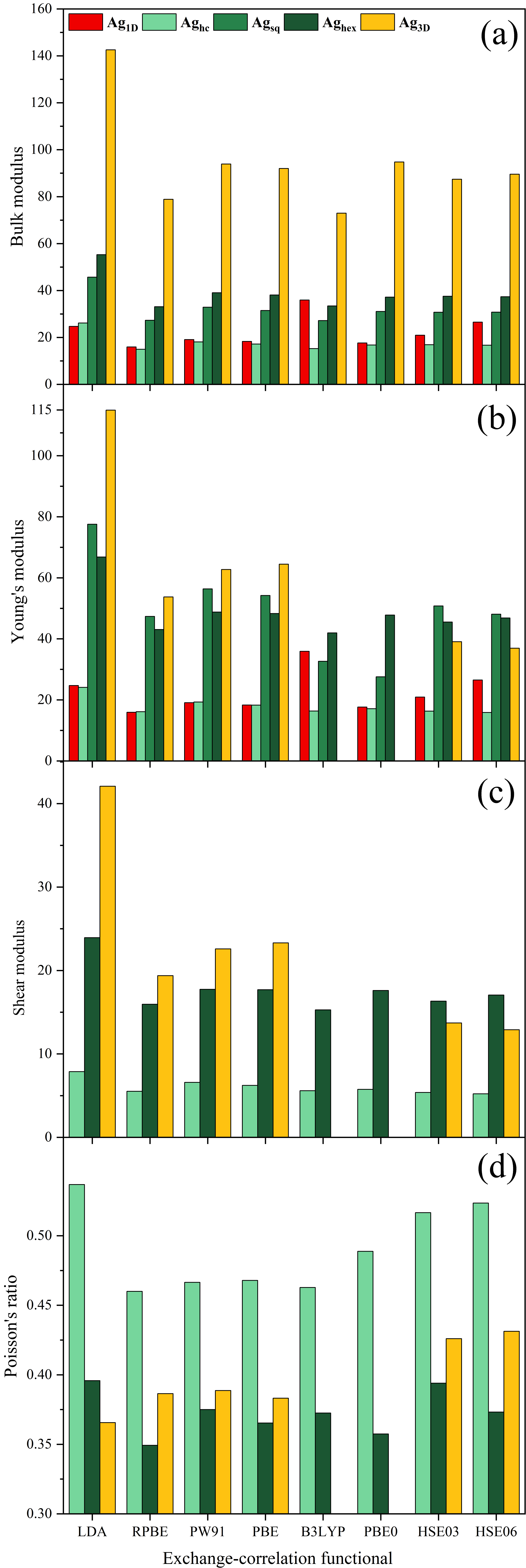}
\caption{Elastic properties of low-dimensional systems of Ag with different xc-functionals. Bulk moduli (a) and Young's moduli (b) are shown for all systems, shear moduli (c) and Poisson's ratio (d) are shown only for 3D and stable 2D systems. Units for moduli are GPa nm$^{3-D}$, where $D$ is the system dimensionality.}
\label{fig:elastic_mod_xc}
\end{figure}

\paragraph{Elastic constants (results):}
Functionals show similar trends for bulk moduli, but there are quantitative differences (Figure \ref{fig:elastic_mod_xc}a). We remind that because the elastic moduli in different dimensions have different units, the trend with respect to the coordination number can be compared only between different 2D lattices. LDA overestimates the bulk moduli systematically, for 3D bulk by almost 40~\%. Only for 1D chain the modulus is in line with HSE06. Among GGAs, the bulk moduli of PW91 and PBE are nearly the same. The hybrid functionals have fairly similar performance, with B3LYP again showing a striking exception, especially related to 1D modulus. These observations in bulk moduli apply also to Young's moduli (Figure \ref{fig:elastic_mod_xc}b). Only GGAs show somewhat larger stiffness and the trends in 2D moduli for B3LYP and PBE0 are different. 

The shear modulus and Poisson's ratio are defined only for 2D and 3D systems (Figures \ref{fig:elastic_mod_xc}c and d). Moreover, shear modulus is not reported for the 2D square lattice due to instability against shear deformations. In addition, some deformations with PBE0 and B3LYP resulted in consistent numerical errors, forcing us to omit shear and Young's modulus as well Poisson ratio for these functionals. In summary, the most consistent behavior in elastic moduli is displayed by HSE and GGA functionals. LDA, B3LYP and PBE0 functionals suffer from both numerical challenges and deviant trends at least in some elastic properties.

\vspace{0.2cm}
 \paragraph{Electronic structure (density of states):}

To complement pure energetic and geometric properties, we now extend our investigations to electronic structure properties. Electronic structure is a complex topic with many features. To reduce complexity and extract trends, we investigate the electronic structure simply in terms of the density of states DOS$(\epsilon)$ and its projections DOS$_l(\epsilon)$ to $s$ ($l=0$), $p$ ($l=1$), and $d$ ($l=2$) angular momentum states. In addition, we focus only on energies at the vicinity of the Fermi-level $\epsilon=\epsilon_F$. 

Consequently, we define the quantities
\begin{equation}
\label{eq:14}
N_{l} = \int^{\infty}_{-\infty} \text{DOS}_{l}\left(\epsilon\right)  g\left(\epsilon\right)\, d\epsilon\, 
\end{equation}
that give the number of $l$-type orbitals surrounding the Fermi-level. The DOS is also normalized by the number of atoms in the simulation cell. The envelope function $g(\epsilon)$ has a Gaussian form
\begin{equation}
\label{eq:15}
g\left(\epsilon\right) = \text{exp}\left[{-\frac{1}{2}\left( \frac{\epsilon - \epsilon_{f}}{\sigma} \right)^2}\right]
\end{equation}
and we used $\sigma=1$~eV energy window around $\epsilon_F$.

\begin{figure}[t]
\centering
\includegraphics[width=8.6cm]{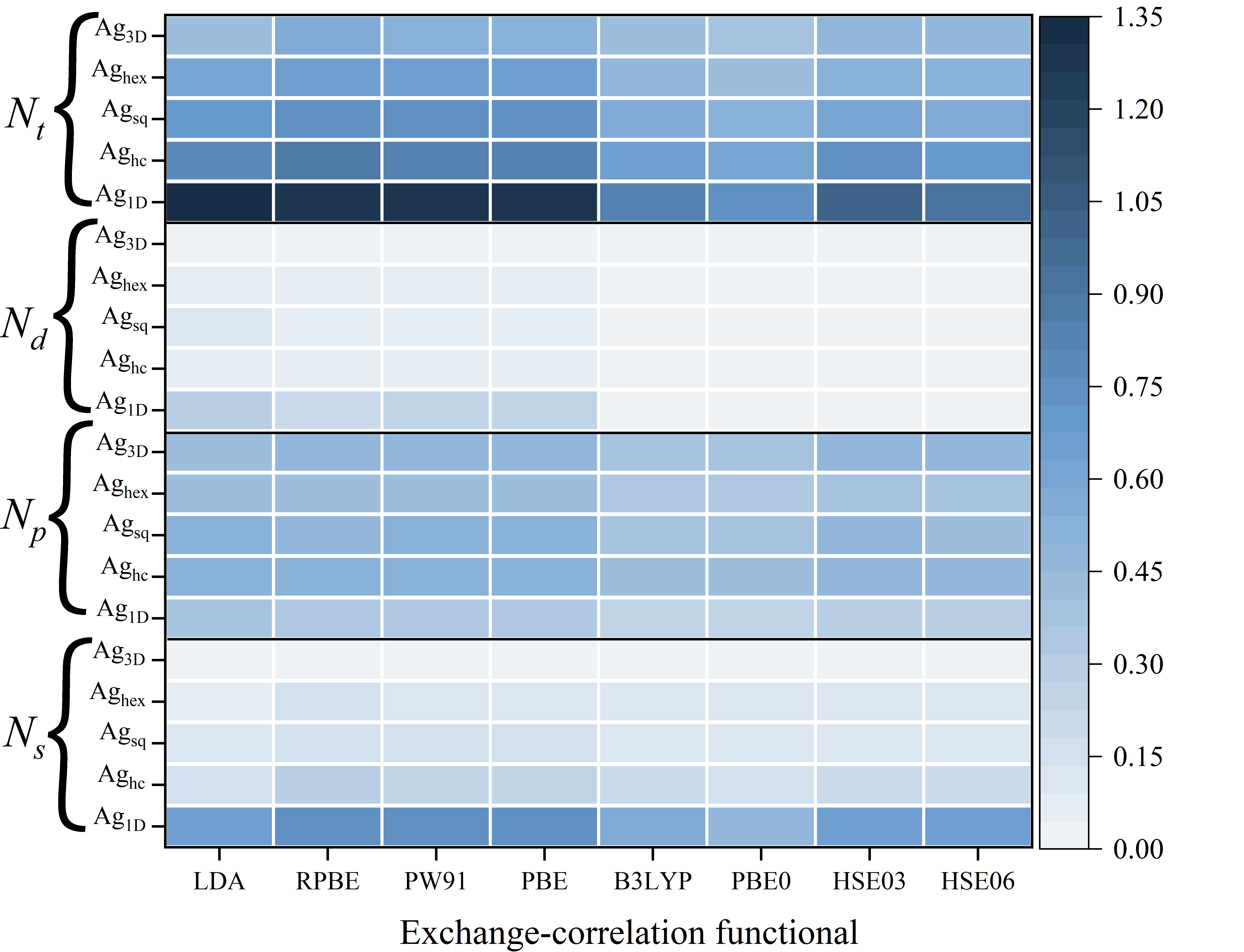}
\caption{Effect of xc functional on the electronic structure of low-dimensional metals made of Ag. Heatmap visualizes the number of $s$-type states ($N_s$), $p$-type states ($N_p$), $d$-type states ($N_d$), and the total number of states ($N_t$) within a $\sim 1$~eV energy window around the Fermi-level [see Eq.(\ref{eq:14})].}
\label{fig:dos_xc}
\end{figure}

In general, the \textit{s}-orbital contribution decreases with increasing coordination number for all xc functionals (Figure~\ref{fig:dos_xc}). In 1D the main contribution comes from \textit{s}-orbitals, followed by \textit{p}- and \textit{d}-orbitals for all functionals. In 2D this order is rearranged to \textit{p} $>$ \textit{s} $>$ \textit{d}. In 3D this same trend is retained by all hybrid functionals. The LDA, PW91, and PBE have very similar orbital contribution ordering. For all xc functionals, the $p$ contribution is the largest for honeycomb, smallest for 1D, and smallest for hexagonal among 2D systems. The ordering of $N_p$ with respect to different coordination number is the same for GGAs, PBE0, and B3LYP. For HSE03 and HSE06 all $N_l$ are very similar. The $d$-orbital contributions follow trend similar to $s$-orbitals. The value of $N_d$ is the highest for LDA and the lowest for PBE0 for all systems; the most visible difference is the generally low $N_d$ of all hybrid functionals, especially in 1D. 

Regarding the total DOS, all GGAs produce nearly identical $N_t$, apart from 3D bulk in RPBE. The total DOS from hybrids differs somewhat from the LDA and GGA functionals. HSE functionals show similar $N_t$ for $C=6$ and $12$ systems, but differ in other systems. Overall, trends in the total densities are inconsistent for LDA and PBE0 functionals, but somewhat consistent among GGA as well as B3LYP and HSE functionals.

\vspace{0.2cm}
\paragraph{Conclusions on xc functionals:}
To summarize, PW91 and PBE perform similarly for forces, energies, and densities of states, while RPBE shows underbinding, smaller bond lengths, and smaller elastic constants. LDA is inferior to GGA practically in all respects. Among hybrid functionals, the performances of HSE03 and HSE06 aligned in all respects. B3LYP failed to improve GGA in terms of accuracy in the lattice constants and cohesive energies, even if its electronic structures resembled those of HSE functionals. Cohesion energy displayed congruent dimensionality-dependencies, apart from visibly differing dependencies by B3LYP and PBE0 functionals. 

Before reaching ultimate conclusions, however, we have to consider the computational cost (Table~\ref{tab:timing_xc}). As expected by the nonlocal character of the hybrid functionals, already minimal-cell systems require $2-3$ orders of magnitude more computational time for hybrids than for LDA and GGA, and for larger systems the difference would increase even further. Considering the low computational cost, GGA functionals perform extremely well compared to hybrid functionals, compared even to the most robust HSE family. To conclude, unless the low-dimensional metals are studied for very specific purposes, the standard PBE indeed remains the preferred weapon of choice for low-dimensional metals modeling.

\begin{table}[h]
\caption{\label{tab:timing_xc}%
Computational cost of different xc-functionals: Time in seconds to calculate the energy of minimal-cell systems using 24 cores. The cell has one atom for all systems except for 2D honeycomb.}
\begin{ruledtabular}
\resizebox{\columnwidth}{!}{
\begin{tabular}{lccccrrrr}
    \multicolumn{1}{l}{}& \multicolumn{1}{c}{LDA} & \multicolumn{1}{c}{RPBE} & \multicolumn{1}{c}{PW91} & \multicolumn{1}{c}{PBE} & \multicolumn{1}{c}{B3LYP} & \multicolumn{1}{c}{PBE0} & \multicolumn{1}{c}{HSE03} & \multicolumn{1}{c}{HSE06} \\
    \hline
        1D   & 39 & 39 & 44 & 43 & 476 & 1360 & 491 & 1897 \\
    hc  & 49 & 59 & 62 & 58 & 16786 & 20937 & 18662 &15006 \\
    sq  & 18 & 24 & 23 & 22 & 1469 & 1739 &1535  &1493 \\
   hex   & 16 & 19  & 20 & 17 & 1454 & 1800 & 1698 & 1675 \\
   3D & 14 & 18 & 19 & 17 & 88553& 41352 & 38802 & 38704 \\
     \end{tabular}}
\end{ruledtabular}
\end{table}
\subsection{\label{sec:basis}Performance of different basis sets}

In this section, we choose PBE xc functional and repeat the systematics of the previous section while this time varying the basis set. The converged plane wave basis gives the best results that provide the reference assessing the performance of the three LCAO basis sets Medium, High, and Ultra introduced in Section~\ref{sec:2}.

To obtain a broader context, we compared the DFT-LCAO with DFTB method, which uses a minimal local basis and contains approximations speeding up the calculations. Here we used the parameters available for Ag developed earlier \cite{DFTB_1,DFTB_2}. However, parametrization can be done in different ways, and one should \emph{not} consider these results as unique and absolute representation of DFTB.

\vspace{0.2cm}
\paragraph{Cohesive Energies:}

\begin{figure}[t!]
\centering
\includegraphics[width=8.6 cm]{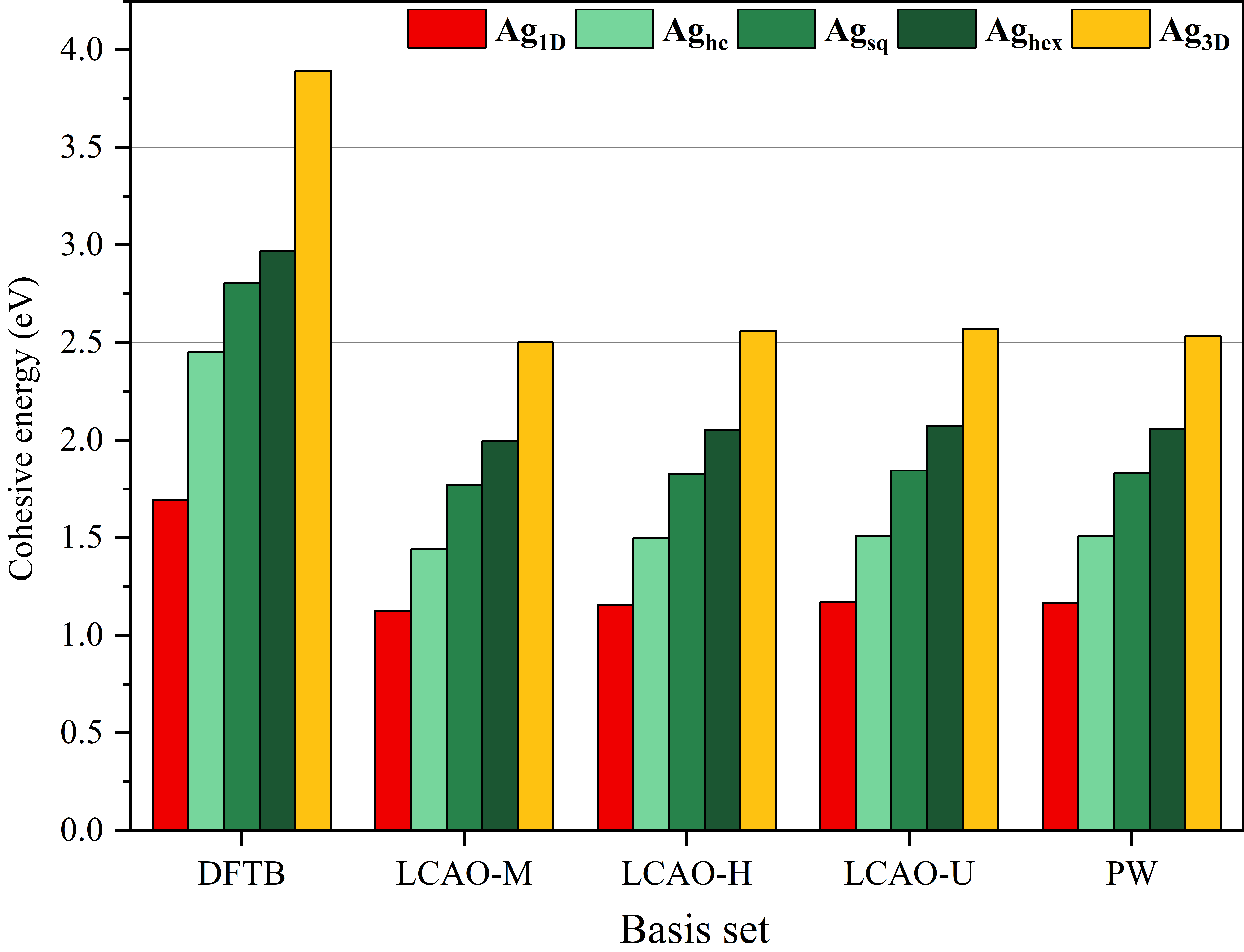}
\caption{Cohesive energies of optimized 1D, 2D (hc, sq, and hex), and 3D systems made of Ag with different basis sets. Bars on the left show DFTB results with minimal basis for comparison.}
\label{fig:coh_basis}
\end{figure}
\vspace{0.2cm}

The LCAO-U and LCAO-H produce cohesive energies very close to those of PW (Figure \ref{fig:coh_basis}). LCAO-M overbinds slightly in comparison, but the accuracy for 2D systems is still $3-4$~\%\ compared to PW. The dependence of cohesion on coordination number is reproduced with all basis sets, and differences are difficult to see on absolute scale. DFTB follows similar behavior, but shows significant overbinding, especially for 3D bulk.

\paragraph{Dimensionality-dependence of energetics:}

As with xc functionals, we investigate how basis set affects the dependence of energetics on coordination number. Again this dependence is analyzed via the scaling exponent $\gamma$ in Eq.~(\ref{eq:gamma}) fitted to the cohesive energies as a function of $C$.

\begin{figure}[t!]
\centering
\includegraphics[width=8.6 cm]{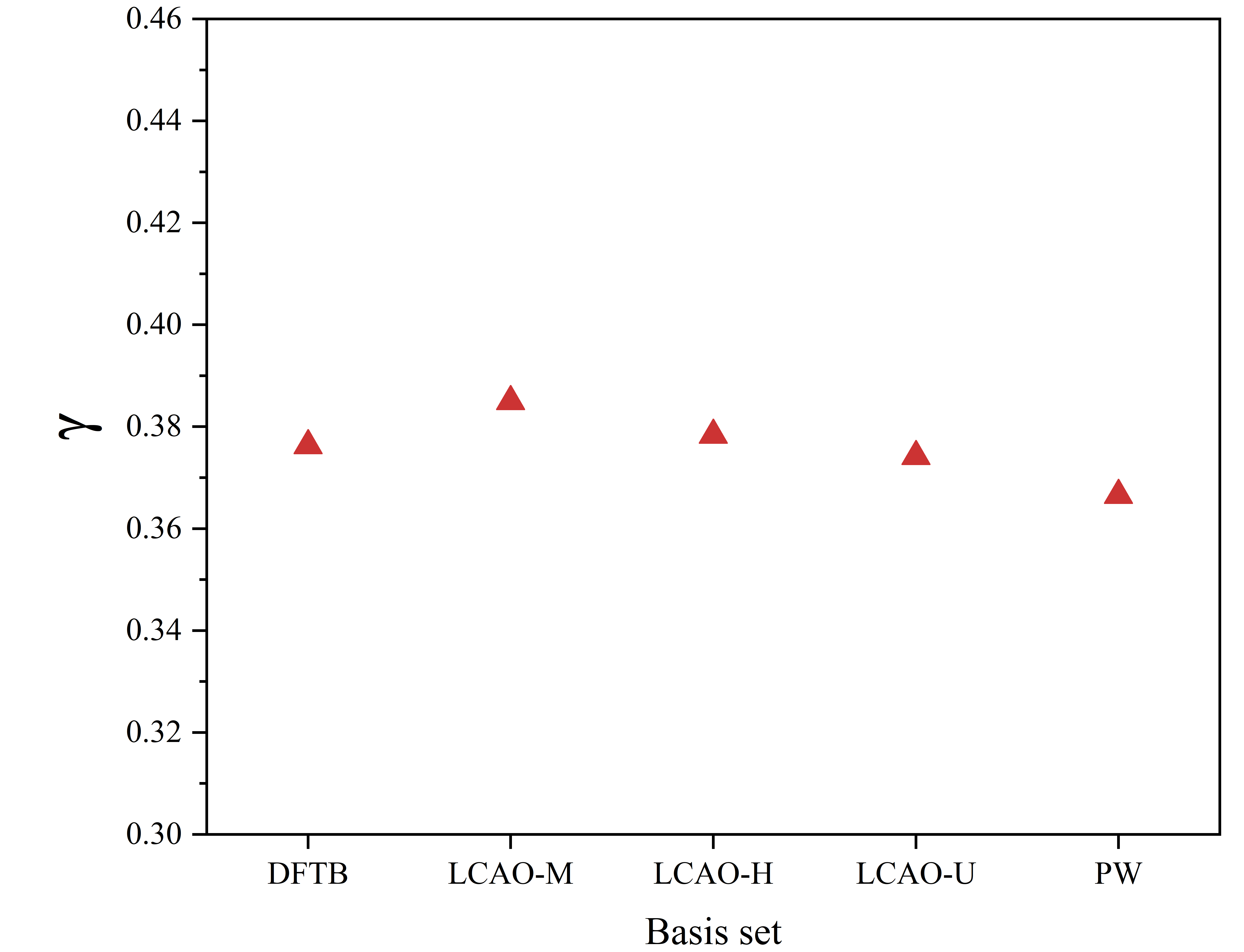}
\caption{Trends of low-dimensional energetics with different basis sets. The fitted scaling exponent $\gamma$ is plotted for different basis sets; smaller $\gamma$ means that energy depends less linearly on the coordination number [see Eq.(\ref{eq:7})]. The vertical scale is the same as in Fig.~\ref{fig:gamm_xc}.}
\label{fig:gamm_basis}
\end{figure}

Compared to PW, the dependence on $C$ becomes systematically more linear as we move from Ultra to High and ultimately to Medium basis (Figure~\ref{fig:gamm_basis}). However, still the \textit{Medium} basis reproduces $\gamma$ to within $5$~\%\ accuracy compared to PW basis. Even DFTB compares well in the overall coordination-dependence, although there are visible problems in capturing the DFT trends for 2D systems (the green bars for DFTB in Figure~\ref{fig:coh_basis}). However, to state the main point, the choice of basis influences dimensionality-dependence of energetics \emph{far less} than xc functional: note that Figs.~\ref{fig:gamm_xc} and \ref{fig:gamm_basis} have the same scale in $\gamma$.

\begin{figure}[t!]
\centering
\includegraphics[width=8.6 cm]{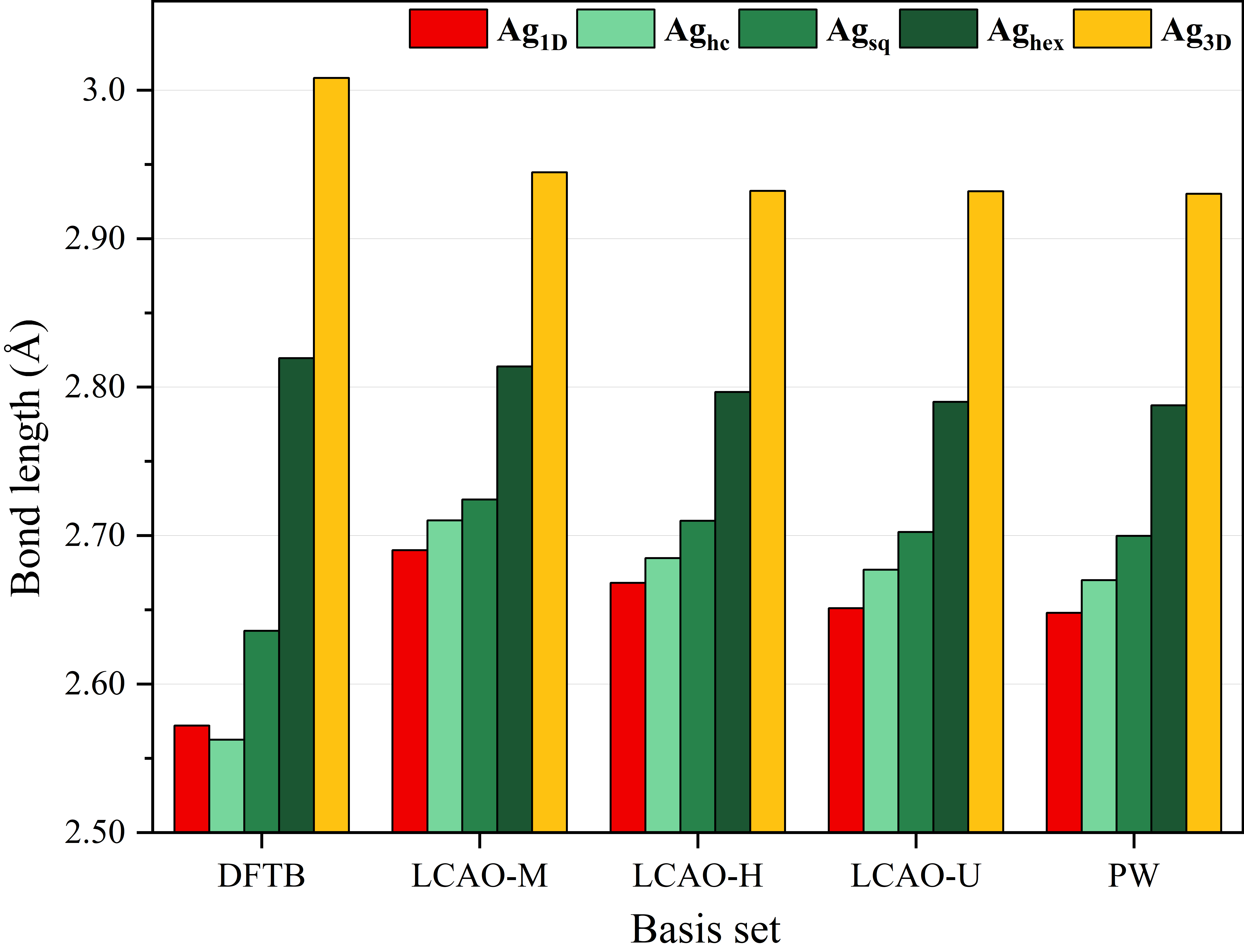}
\caption{Bond lengths of optimized 1D, 2D (hc, sq, and hex), and 3D systems made of Ag with different basis sets.}
\label{fig:bond_basis}
\end{figure}

\vspace{0.2cm}
\paragraph{Bond Lengths:}
The LCAO-U and LCAO-H bond lengths are very similar, accurate to within $0.77$~\%\ compared to PW (Figure \ref{fig:bond_basis}). All LCAO variants overestimate all bonds, LCAO-M having the lowest performance with $1.6$~\%\ too long bonds. DFTB no longer captures the DFT trends in coordination-dependence. The 1D chain bond length is larger than honeycomb and the 2D bonds vary wildly, even if the $C$-ordering still remains correct.

\vspace{0.2cm}

\paragraph{Elastic constants and moduli:}

\begin{figure}
\centering
\includegraphics[width=8.6cm, height = 22.1cm]{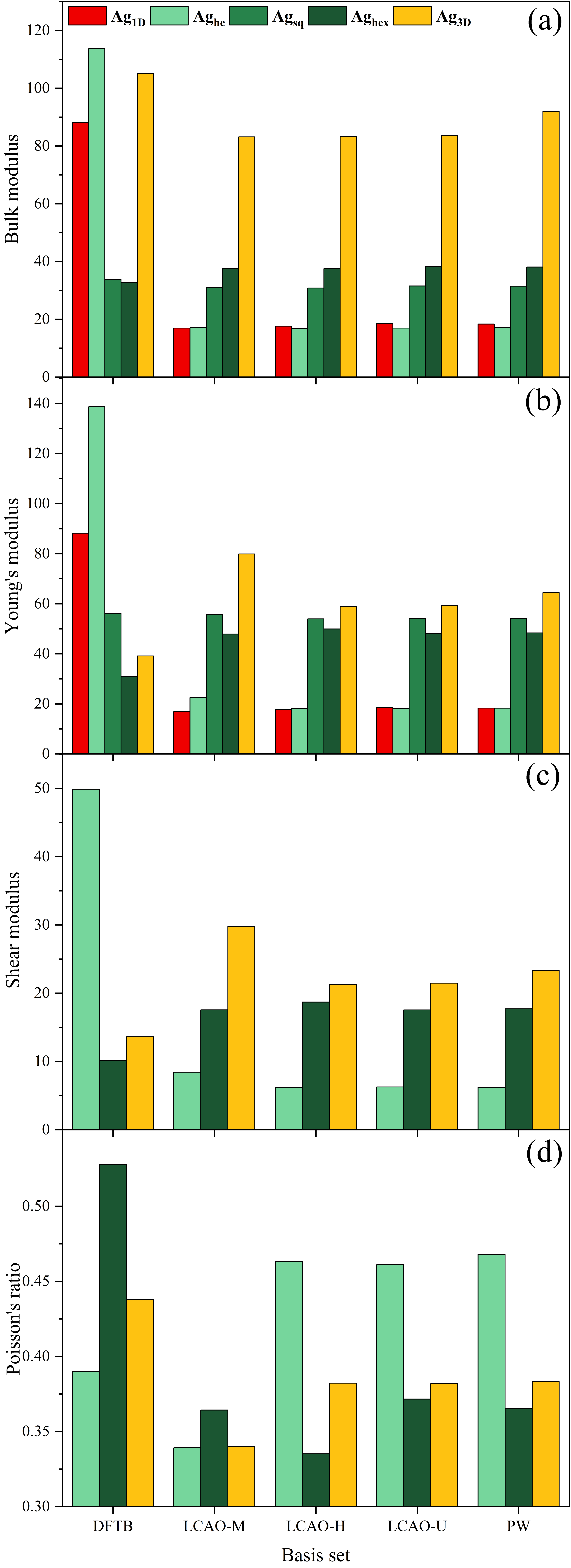}
\caption{Elastic properties of low-dimensional systems of Ag with different basis sets. Bulk moduli (a) and Young's moduli (b) are shown for all systems, shear moduli (c) and Poisson's ratio (d) are shown only for 3D and stable 2D systems. Units for moduli are GPa nm$^{3-D}$, where $D$ is the system dimensionality.}
\label{fig:elastic_mod_basis}
\end{figure}

For 1D and 2D systems, elastic moduli have minor dependence on basis set (Figure \ref{fig:elastic_mod_basis}). The largest deviation from PW occurs for 3D bulk, for all LCAO variants. This deviation likely stems from the better space-filling character of PW basis. Moreover, although performing well in cohesion and bond lengths, LCAO-M performs poorly in all elastic properties. LCAO-U is close to PW in all respects, and LCAO-M captures all the same trends, even if with some quantitative differences. These results suggest that, except perhaps for LCAO-M, LCAO basis can be reliable for studying mechanical properties of low-dimensional metallic systems. The LCAO variant -dependency of elastic properties is even smaller than the changes upon switching from GGA to hybrid functional (compare Figs. \ref{fig:elastic_mod_xc} and \ref{fig:elastic_mod_basis}). 

In comparison, DFTB shows both trend differences and large absolute differences compared to DFT-LCAO (Figure~\ref{fig:elastic_mod_basis}). For example, the 1D elastic modulus is overestimated by a factor of $\sim 5$. Even the trend within 2D systems was not reproduced. It appears that the Ag parametrization should be revised for more reliable mechanical properties of low-dimensional Ag systems.

\vspace{0.2cm}

\paragraph{Electronic structure (density of states):}

\begin{figure}[t!]
\centering
\includegraphics[width=8.6 cm, height=7cm]{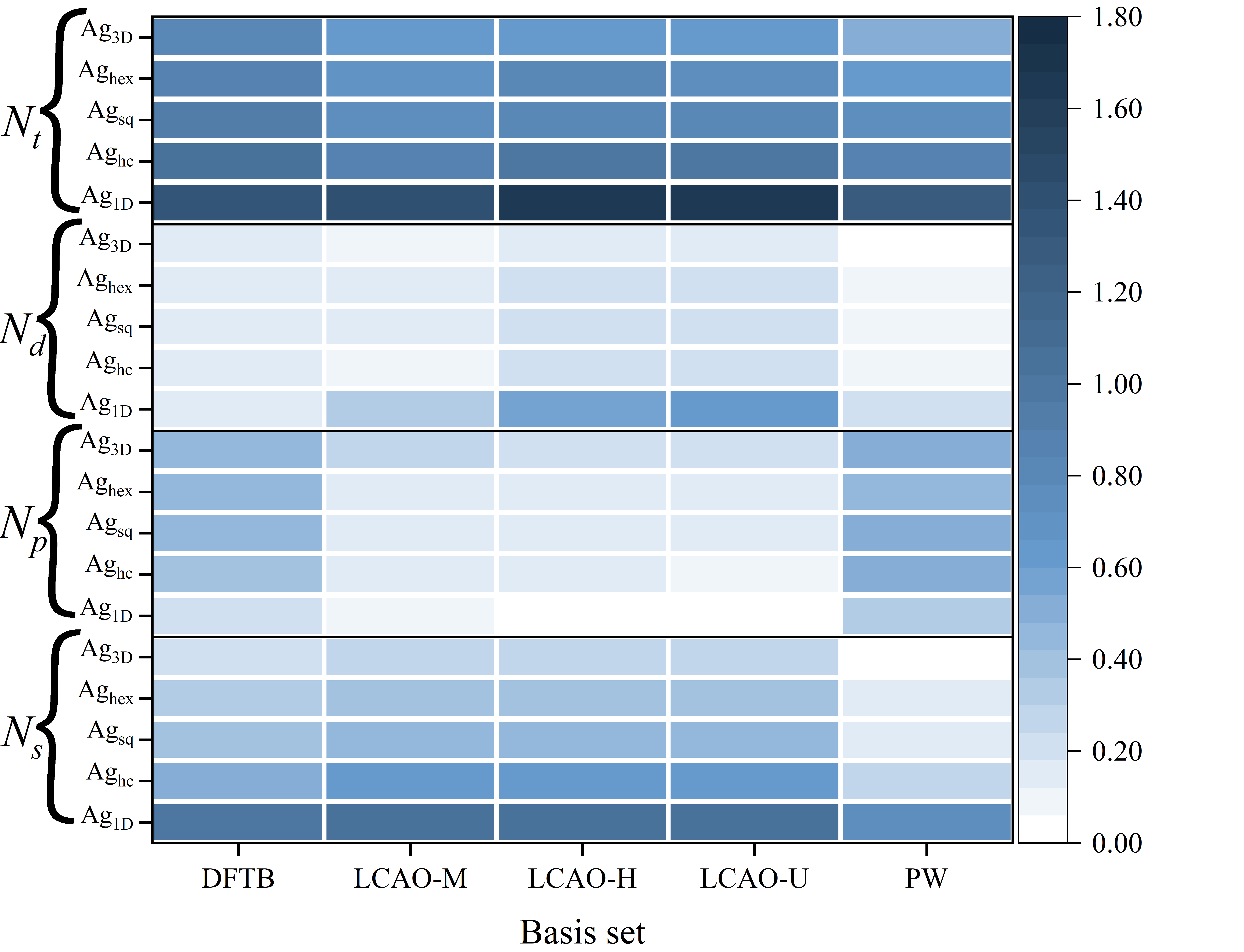}
\caption{Effect of basis set on the electronic structure of low-dimensional metals made of Ag. Heatmap visualizes the number of $s$-type states ($N_s$), $p$-type states ($N_p$), $d$-type states ($N_d$), and total number of states ($N_t$) within a $\sim 1$~eV energy window around the Fermi-level [see Eq.(\ref{eq:15})].}
\label{fig:dos_basis}
\end{figure}

Also the electronic structure from LCAO is compared here against PW results, using the indicator numbers given by Eq.~(\ref{eq:14}). For 2D structures PW gives orbital contributions in order $p>s>d$ (Figure \ref{fig:dos_basis}). For LCAO this trend shuffles to $s>d>p$, that is, the $p$ contribution diminishes for all LCAO variants. For 1D system the orbital ordering for PW and LCAO basis remains the same. However, still all basis sets---including minimal-basis DFTB---show consistent $C$-dependence in orbital contributions around the Fermi-level. LCAO-H and LCAO-U results align better, while LCAO-M results are different in some respects. In summary, the $C$-dependence of the total DOS in 2D metals is reproduced by LCAO to a fair degree, but the orbital contributions are different.

\paragraph{Conclusions on basis sets:}

To conclude, LCAO basis competes extremely well with PW for studying energetic and geometric properties of low-dimensional metal systems. Even elastic moduli are reproduced reasonably well by LCAO-H and LCAO-U basis, compared to converged PW basis. The performance of LCAO-M basis was notably modest, regarding elastic properties and also the details of electronic structure. The orbital breakup of the electronic structures at the vicinity of Fermi-level for PW and LCAO variants differed markedly.

Regarding DFTB, the Ag parametrizations clearly require revisiting. The cohesive energies are too large, bond lengths are both large and small, and elastic moduli are close to arbitrary. Still many of the qualitative trends regarding $C$-dependence were reproduced reliably.

However, before again reaching ultimate conclusions, we have to consider the computational cost with different basis (Table \ref{tab:time_basis}). The cost was investigated by simulation cells with $32-64$ atoms and a couple of dozen cores. The comparison is thus by no means unique or absolute, but it does give a rough inkling of the computational demands. As expected, DFTB outspeeds DFT by one to three orders of magnitude. Within DFT, switching from LCAO-M to LCAO-U results in cost increases from a factor of two (1D) up to a factor of $\sim 15$ (3D). Especially for low-dimensional systems LCAOs are faster than PW, nearly by two orders of magnitude. For 3D bulk PW is very competitive against LCAO due to lacking vacuum region; here LCAO-U is even slower than PW. In conclusion, unless very high accuracy is of central importance, LCAO has demonstrated a fair accuracy in most properties and should be prioritized over PW due to its superior efficiency. Even LCAO-M basis can be considered for simulations where the improved speed wins over lost accuracy.

\begin{table}[t!]
\caption{\label{tab:time_basis}%
Computational cost of different basis sets: Time in seconds to calculate the energy of systems using 24 cores. The parenthesis contain the number of atoms in the supercell.
}
\begin{ruledtabular}
\begin{tabular}{lccccc}
    \multicolumn{1}{l}{Systems} & \multicolumn{1}{c}{DFTB} & \multicolumn{1}{c}{LCAO-M} & \multicolumn{1}{c}{LCAO-H} & \multicolumn{1}{c}{LCAO-U} & \multicolumn{1}{c}{PW} \\
    \hline    
    1D (32) & 10 & 175  & 265 & 310 & 11890 \\
    2D hc (64) & 20 & 215 & 355 & 610 & 13120 \\
    2D sq (64) & 18    & 190 & 300 & 500 & 12370 \\
    2D hex(64) & 17 & 130 & 290 & 655 & 6885 \\
    3D (64) & 19  & 145   & 855  & 2220  & 2050 \\     
\end{tabular}
\end{ruledtabular}
\end{table}

\subsection{\label{sec:citeref}Combined scanning of xc functionals and basis sets}

Above we investigated xc functionals (with PW basis) and basis sets (with PBE functional) separately. However, the performance of xc functionals and basis sets can be coupled. We therefore complement our analysis by combined scanning of different xc functionals with different basis sets. The bond lengths, cohesive energies, elastic constants, and orbital contributions to DOS obtained at different basis set-xc functional -combinations are shown in Tables V, VI, and VII in the Appendix.

For LDA, the choice of basis set did not affect the cohesion dependence on $C$ (Table V). Changing the basis set from PW to LCAO increases the cohesive energy for $C\geq 4$ and decreases it for $C=1$ and $3$. Decreasing the LCAO size also decreases the cohesion, as expected in the light of variational principle. Bond lengths with PW, LCAO-U and LCAO-H basis are nearly equal. With LCAO-M bonds are longer for all systems. The elastic properties are nearly basis-independent, with the notable exception of LCAO-M (Table VI). Most sensitive to the choice of basis is the electronic structure; all LCAO variants show the same trend, which however differs significantly from PW (Table VII). 

For GGAs, the performance remains robust upon reducing the size of the basis set. In fact, the observations in Subsection~\ref{sec:basis} with PBE are representative for other GGAs as well. Switching PW to LCAO-U or LCAO-H changes bond lengths and cohesive energies less than $1$~\%; less robust LCAO-M decreases cohesive energies by $4$~\%\ and increases bond lengths by $\approx 1.5$~\%\ (Table V). Basis set sensitivity is the smallest for PW91 and the largest for RPBE. Elastic constants follow the accuracy trends similar to those of energetics and geometric properties. PBE shows some basis set sensitivity, especially for the bulk moduli of 2D systems (Table VI).

For hybrid functionals, the matters are less systematic. Using LCAO-M in conjunction with unscreened B3LYP and PBE0 functionals results in significant overbinding; bond lengths are underestimated by more than $10$~\% (Table V). With LCAO-H and LCAO-U basis sets the same xc functionals underestimate bonds only by $\approx 2$~\%, while increase cohesive energies by $\leq 24$~\%. B3LYP and PBE0 are thus extremely sensitive to the quality of LCAO basis. Moreover, B3LYP and PBE0 are unable to produce elastic moduli due to persistent numerical errors. In contrast, the screened HSE functionals produced robust geometries, energetics and elastic properties upon changing the size of the LCAO basis. The robustness was even better than with PW91 and PBE, although admittedly at a considerable computational cost. The orbital contributions to DOS with PW and LCAO basis were different; the same effect was observed for PBE functional (Figure~\ref{fig:dos_basis}). Among different LCAO variants, LCAO-H and LCAO-U show similar orbital contributions for all systems. In addition to energetic and geometric properties, the peculiarities of B3LYP and PBE0 functionals are observable also in electronic properties (Table VII). In general, hybrid functionals in conjunction with LCAO-H and LCAO-U basis requires prohibitive computational resources even for single atom.

	\subsection{\label{sec:codes}The effect of DFT implementation}
	In addition to DFT attributes, it is important also to be able to rely on the DFT implementation itself. For completeness, therefore, we briefly discuss the magnitude of differences related to the numerical implementation of DFT. We calculated the cohesive energies, bond lengths, and elastic moduli also with the GPAW code, using plane wave basis with the same $800$~eV energy cutoff and default parameters \cite{GPAW}. The QuantumATK/GPAW cohesive energies were $1.1671$\,eV / $1.1661$\,eV (1D), $1.5062$\,eV / $1.5054$\,eV (2D hc), $1.8293$\,eV / $1.8286$\,eV (2D sq), $2.0583$\,eV / $2.0570$\,eV (2D hex), $2.5326$\,eV / $2.5323$\,eV (3D), bond lenghts $2.6480$\,\AA / $2.6501$\,\AA\ (1D), $2.6700$\,\AA / $2.6682$\,\AA\ (2D hc), $2.6998$\,\AA / $2.700567$\,\AA\ (2D sq),  $2.7877$\,\AA / $2.7894$\,\AA\ (2D hex), $2.9301$\,\AA / $2.9305$\,\AA\ (3D), and bulk moduli $18.32$\,GPa nm$^2$ /$18.73$\,GPa\,nm$^2$ (1D), $17.20$\,GPa\,nm / $17.21$\,GPa\,nm (2D hc), $31.46$\,GPa\,nm / $31.26$\,GPa\,nm (2D sq), $38.07$\,GPa\,nm / $37.79$\,GPa\,nm (2D hex), $92.03$\,GPa / $90.37$\,GPa (3D). Thus, default parameters without tuning give code-related differences in cohesive energies $\lesssim 1.3$~meV, in bond lengths $\lesssim 0.002$~\AA, and in bulk moduli $\lesssim 1$~\% (2D systems) or $\simeq$ 2\% (1D and 3D systems). Although the comparison used the PBE functional and plane waves, it is reasonable to suspect the level of differences to remain similar also for other functionals and basis sets. Overall, code-related differences remain considerably smaller than the differences originating from physical attributes.

\section{\label{sec:level4}Summary and Conclusion}

In summary, we investigated the performance of various DFT attributes in the modeling of low-dimensional elemental metals. For future reference, the number of k-points, the size of the vacuum region, and the magnitude of Fermi-broadening were given tolerance-dependent rules of thumb. Such rules help choosing combinations of attributes that result in commensurate accuracies.

The most robust against the choice of basis set was HSE06, followed by HSE03, PBE, PW91, RPBE and LDA. The B3LYP produced inaccurate cohesions and bond lengths---with the highest computational cost. Only the electronic structure in B3LYP was in line with other hybrid functionals. 

The energetics, geometries, and elastic properties with PW, LCAO-U, and LCAO-H basis sets were in overall good agreement. The greatest disparities between PW and LCAO methods resided in the orbital contributions to the DOS, although in the total DOS they were moderated. On a general level, LCAO-U and LCAO-H performed similarly at different xc functionals; therefore, for general purposes, LCAO-H should be preferred over LCAO-U due to superior efficiency (Table~\ref{tab:time_basis}). The LCAO-M basis worked varyingly well in many respects, except when used in conjunction with B3LYP and PBE0 functionals. 

To conclude, in the research of metallic bonding at low dimensions, the best value for a given cost is probably given by semi-local PW91 and PBE xc functionals in conjunction with moderately-sized LCAO-U or LCAO-H basis sets. These results are encouraging for doing large-scale, high-throughput DFT simulations to generate data for machine learning algorithms. In comparison, DFTB is a very speedy method and is capable of simulations unaccessible by DFT \cite{plenty,Liquid,Koskinen_2006}, but the quality of parametrization needs to be ensured first. We hope that our results and gentle recommendations help lifting 2D metal research to new heights, expedite better interaction with experiments, and feed machine learning algorithms with quality data to drive further discoveries in low-dimensional metals.
 
\begin{acknowledgments}
We acknowledge the Finnish Grid and Cloud Infrastructure (FGCI) for computational resources.
\end{acknowledgments}


%

\begin{table*}[h!]
    \centering
     $\bf APPENDIX$\\
     \caption{\label{tab:bond}Bond lengths \(d(\si{\angstrom})\) and Cohesive energies \(E_{coh}\)(eV) for each lattice type corresponding to different DFT-attributes.}
   \begin{ruledtabular}
\begin{tabular}{l|cccccccccc}
        &\multicolumn{2}{c}{1D}&\multicolumn{2}{c}{Honeycomb} &\multicolumn{2}{c}{Square}&\multicolumn{2}{c}{Hexagonal}&\multicolumn{2}{c}{3D}\\ \hline
 \\
DFT-Methods&\(d\)&\(E_{coh}\)&\(d\)&\(E_{coh}\)&\(d\)&\(E_{coh}\)&\(d\)&\(E_{coh}\)&\(d\)&\(E_{coh}\)
\\ \hline
DFTB &2.572&1.691&2.562 &2.450&2.636&2.804&2.819&	2.967&	3.008&		3.891\\
LDA-LCAO-M&	2.584&	1.513	&2.591&	2.012&	2.623	&2.475&	2.712&	2.761&	2.840&	3.547\\
LDA-LCAO-H&	2.553&	1.563&	2.562&	2.105&	2.606&	2.563&	2.693&	2.858&	2.827&	3.660\\
LDA-LCAO-U&		2.542&	1.587&	2.553&	2.126&	2.598&	2.590&	2.685&	2.887&	2.826&	3.672\\
LDA-PW	&2.542&	1.591&	2.542	&2.138	&2.595	&2.586&	2.682&	2.881&	2.828&	3.638\\
RPBE-LCAO-M&2.732&	0.959	&2.760&	1.198	&2.764&	1.474&	2.853	&1.677&	2.982	&2.065\\
RPBE-LCAO-H&2.710&	0.989	&2.731	&1.251	&2.745&	1.531	&2.831	&1.738	&2.965	&2.130\\
RPBE-LCAO-U&2.691&	1.001	&2.723	&1.262&	2.736&	1.547	&2.824	&1.756	&2.963	&2.143\\
RPBE-PW	&2.689	&0.992	&2.709	&1.248	&2.734	&1.523&	2.822&	1.732	&2.962&	2.100\\
PW91-LCAO-M&	2.679&	1.145&	2.700	&1.470	&2.717	&1.806&	2.807	&2.026	&2.941&	2.529\\
PW91-LCAO-H&	2.655&	1.171&	2.670&	1.522&	2.703&	1.858&	2.790&	2.083&	2.932&	2.586\\
PW91-LCAO-U&	2.642&	1.186	&2.668&	1.536&	2.696&	1.876&	2.785&	2.103&	2.932&	2.598\\
PW91-PW&	2.639&	1.185&	2.659&	1.534&	2.693&	1.862&	2.783&	2.089&	2.928&	2.560\\
PBE-LCAO-M&2.690	&1.126	&2.710	&1.441	&2.724&	1.771&	2.814	&1.994&	2.945&	2.501\\
PBE-LCAO-H&2.668	&1.155	&2.685&	1.497&	2.710&	1.826&	2.797&	2.053&	2.932	&2.558\\
PBE-LCAO-U&2.651	&1.170	&2.677	&1.510	&2.702&	1.844&	2.790&	2.073	&2.932&	2.571\\
PBE-PW&	2.648&	1.167&	2.670&	1.506	&2.700	&1.829&	2.788&	2.058&	2.930&	2.533\\
B3LYP-LCAO-M&	2.373&	3.734&	2.410&	5.164&	2.457&	6.029&	2.558&	6.586&	2.725&	8.340\\
B3LYP-LCAO-H	&2.655&	1.067&	2.691&	1.426&	2.714&	1.772	&2.812&	1.977&	-&	-\\
B3LYP-LCAO-U	&2.642	&1.100&	2.679	&1.461	&2.705&	1.816	&2.803	&2.025	&-&	-\\
B3LYP-PW&	2.681	&0.944&	2.715&	1.211&	2.737	&1.470	&2.830	&1.659&	2.986&	1.963\\
PBE0-LCAO-M&2.322&	4.877&	2.358&	6.807	&2.409&	7.978	&2.512	&8.657&	-&	-\\
PBE0-LCAO-H&2.635&	1.092&	2.654&	1.523	&2.679&	1.970	&2.773	&2.219&	-&	-\\
PBE0-LCAO-U&2.626&	1.128&	2.642&	1.567	&2.670&	2.023	&2.764	&2.277&	-	&-\\
PBE0-PW&2.649	&0.963&	2.671&	1.288	&2.690&	1.640	&2.779	&1.879	&2.910	&2.444\\
HSE03-LCAO-M&	2.694&	1.030	&2.715&	1.351&	2.729&	1.696	&2.825	&1.919&	2.725&	2.436\\
HSE03-LCAO-H&2.668&	1.044	&2.693	&1.385	&2.714	&1.728	&2.807&	1.949&	-&	-\\
HSE03-LCAO-U&2.663&	1.058	&2.687	&1.396	&2.710	&1.744	&2.801	&1.966&	-&	-\\
HSE03-PW&2.651&	1.049	&2.664	&1.392&	2.697	&1.742	&2.787	&1.971&	2.925&	2.484\\
HSE06-LCAO-M&	2.697&	1.061&	2.716&	1.358&	2.733&	1.707&	2.829	&1.932	&2.954	&2.431\\
HSE06-LCAO-H&	2.676&	1.075&	2.693	&1.391	&2.719	&1.738&	2.812	&1.961	&-&	-\\
HSE06-LCAO-U&	2.666&	1.088&	2.686	&1.402&	2.709&	1.753&	2.803&	1.978&	-	&-\\
HSE06-PW&2.650	&1.075&	2.664&	1.396	&2.695	&1.750	&2.786	&1.982&2.923	&2.479\\
    \end{tabular}
    \end{ruledtabular}
\end{table*}

 \begin{table*}[]
     \centering
     \caption{\label{tab:elastic}Elastic constants for 1D (GPa nm$^{2}$), 2D (GPa nm), and 3D (GPa) calculated  by using different DFT-attributes.}
     \begin{ruledtabular}
   \begin{tabular}{l|ccccccccccccc}
     &\multicolumn{1}{c}{1D}&\multicolumn{3}{c}{Honeycomb} &\multicolumn{3}{c}{Square}&\multicolumn{3}{c}{Hexagonal}&\multicolumn{3}{c}{3D}\\ \hline
 \\
 DFT-Methods&\(C_{11}\)&\(C_{11}\)&\(C_{12}\)&\(C_{66}\)&\(C_{11}\)&\(C_{12}\)&\(C_{66}\)&\(C_{11}\)&\(C_{12}\)&\(C_{66}\)&\(C_{11}\)&\(C_{12}\)&\(C_{66}\)
 \\ \hline
 DFTB&	88.2&	163.6	&63.8&	49.9&	57.8&	9.7	&-3.9&	42.7&	22.6&	10.1&	110.7&	102.4&	19.9\\
 LDA-LCAO-M&	24.5&	34.3&	14.9&	9.7&	80.5&	9.0&	-5.9&	77.9&	30.7&	23.6&	163.6&	133.0&	53.4\\ 
LDA-LCAO-H&	25.2&	33.7&	17.0&	8.3&	79.5&	10.7&	-7.5&	79.1&	28.4&	25.3&	165.4&	131.0&	56.3\\
LDA-LCAO-U	&25.8&	34.2&	17.4&	8.4&	80.6&	12.1&	-7.6&	85.3&	27.1&	29.1&	164.3&	131.4&	54.7\\
LDA-PW&	24.7&	34.0&	18.3&	7.9&	79.4&	12.0&	-8.8&	79.2&	31.4&	23.9&	165.4&	131.1&	58.7\\
RPBE-LCAO-M&	15.5&	19.0&	8.7	&5.1	&48.6&	4.6&	-2.9&	43.5&	13.8&	14.9&	103.4&	88.0	&35.9\\

RPBE-LCAO-H&	15.3&	19.4	&9.0&	5.2&	47.8&	5.6&	-2.3&	48.6&	16.7&	16.0&	103.6&	83.9&	32.7\\
RPBE-LCAO-U&	15.1&	19.6&	8.4&	5.6&	47.9&	6.5&	-2.7&	44.6&	21.8&	11.4&	103.4&	82.9&	31.3\\
RPBE-PW&	16.0&	20.5&	9.4&	5.5&	48.2&	6.4&	-3.4&	49.0&	17.1&	16.0&	92.7&	72.0&	25.4\\
PW91-LCAO-M	&18.8&	24.1&	10.7&	6.7&	57.3&	6.0&	-3.0&	56.2&	-9.0&	32.6&	133.3&	81.2&	16.7\\
PW91-LCAO-H	&18.6&	24.5&	11.7&	6.6&	56.7&	7.4&	-3.4&	56.3&	21.8&	17.2&	116.4&	69.2&	19.7\\
PW91-LCAO-U	&19.1&	24.2&	11.0&	6.6&	56.1&	8.1&	-3.6&	56.8&	21.1&	17.8&	117.7&	68.9&	19.0\\
PW91-PW	&19.1&	24.7&	11.5&	6.6&	57.5&	8.2&	-4.1&	56.8&	21.3&	17.7&	109.8&	85.9&	29.7\\
PBE-LCAO-M&	16.9&	25.5&	8.63&	8.4&	56.2&	5.5&	-3.1&	55.2&	20.1&	17.5&	114.1&	67.6&	34.2\\
PBE-LCAO-H&	17.6&	23.0&	10.7&	6.2&	54.8&	6.8&	-3.6&	56.2&	18.8&	18.7&	113.2&	68.3&	20.5\\
PBE-LCAO-U&	18.6&	23.2&	10.7&	6.2&	55.3&	7.7&	-3.7&	55.8&	20.7&	17.5&	115.2&	68.0&	20.0\\
     \end{tabular}
    \end{ruledtabular}
 \end{table*}

\begin{table*}[h!]
    \centering
TABLE VI. (\textit{Continued})\\
\begin{ruledtabular}
\begin{tabular}{l|ccccccccccccc}
     &\multicolumn{1}{c}{1D}&\multicolumn{3}{c}{Honeycomb} &\multicolumn{3}{c}{Square}&\multicolumn{3}{c}{Hexagonal}&\multicolumn{3}{c}{3D}\\ \hline
 \\
 DFT-Methods&\(C_{11}\)&\(C_{11}\)&\(C_{12}\)&\(C_{66}\)&\(C_{11}\)&\(C_{12}\)&\(C_{66}\)&\(C_{11}\)&\(C_{12}\)&\(C_{66}\)&\(C_{11}\)&\(C_{12}\)&\(C_{66}\)
 \\ \hline
PBE-PW&	18.3&	23.4&	11.0&	6.2&	55.3&	7.7&	-4.3&	55.8&	20.4&	17.7&	107.7&	84.2&	31.0\\
B3LYP-LCAO-M&	65.3&	97.4&	45.6&	25.9&	160.2&	52.2&	-31.9&	168.8&	85.3	&41.8&	-&	-&	-\\
B3LYP-LCAO-H&	19.4&	23.3&	10.5&	6.4&	44.9&	17.0&	-3.6&	51.3&	19.2&	16.0&	-&	-&	-\\
B3LYP-LCAO-U&	20.5&	24.2&	10.7&	6.7&	46.0&	18.2&	-3.4&	52.8&	20.6&	16.1&	-	&-&	-\\
B3LYP-PW&	35.9&	20.8&	9.6&	5.6&	38.9&	15.5&	-1.8&	47.2&	17.6&	14.8&	-&	-&	-\\
PBE0-LCAO-M&	80.4&	115.1&	58.9&	28.1&	205.4&	60.0&	-90.1&	203.9&	103.2&	50.3&	-&	-&	-\\
PBE0-LCAO-H&	20.2&	25.1&	11.1&	7.0&	48.4&	23.8&	-8.0&	58.3&	20.2&	19.1&	-&	-&	-\\
PBE0-LCAO-U&	20.8&	26.3&	14.8&	5.8&	45.2&	25.2&	-8.3&	59.9&	21.9&	19.0&	-&	-&	-\\
PBE0-PW	&17.6&	22.5&	11.0	&5.7&	40.0&	22.2&	-5.3&	55.1&	19.6&	17.6&	138.4&	73.0&	-\\
HSE03-LCAO-M&	17.7	&23.1&	10.2&	6.5&53.9&	8.3&	-3.1&	54.0&	19.6&	17.2&	96.4&	84.5&	27.9\\
HSE03-LCAO-H&	18.0&	23.4&	10.6&	6.4&	50.9&	10.2&	-3.3&	53.2&	20.5&	16.3&	-&	-&	-\\

HSE03-LCAO-U&	17.4&	22.7&	10.9&	5.9&	50.3	&10.0&	-3.5&	53.8&	19.4&	17.2&	-&	-&	-\\
HSE03-PW&	20.9&	22.3&	11.5&	5.4&	52.4	&9.1&	-4.5&	53.8&	21.2&	16.3&	96.2&	83.0&	13.7\\
HSE06-LCAO-M&	17.4&	23.1&	10.3&	6.4&	52.1&	8.6&	-3.1&	52.8&	19.3&	16.8&	113.5&	95.6&	36.5\\
HSE06-LCAO-H&	18.6&	23.2&	10.6&	6.3&	49.9	&11.0&	-3.2&	51.9&	19.5&	16.2&	-&	-&	-\\
HSE06-LCAO-U&	17.1&	24.0&	9.9&	7.0&	49.1&	11.3&	-3.5&	53.3&	18.9&	17.2&	-&	-&	-\\
HSE06-PW&	26.5&	21.9&	11.5&	5.2&	50.5&	11.1&	-5.3&	54.4&	20.3&	17.0&	94.2&	87.2&	14.4\\
    \end{tabular}
    \end{ruledtabular}
\end{table*}

 \begin{table*}[]
     \centering
  \caption{\label{tab:pdos} Estimation of contribution of \textit{s}, \textit{p}, and \textit{d} orbitals to the density of states by implementing different DFT-attributes}
\begin{ruledtabular}
\begin{tabular}{l|ccccccccccccccc}
 &\multicolumn{3}{c}{1D}&\multicolumn{3}{c}{Honeycomb} &\multicolumn{3}{c}{Square}&\multicolumn{3}{c}{Hexagonal}&\multicolumn{3}{c}{3D}\\ \hline
 \\
 DFT-Methods&\(N_{s}\)&\(N_{p}\)&\(N_{d}\)&\(N_{s}\)&\(N_{p}\)&\(N_{d}\)&\(N_{s}\)&\(N_{p}\)&\(N_{d}\)&\(N_{s}\)&\(N_{p}\)&\(N_{d}\)&\(N_{s}\)&\(N_{p}\)&\(N_{d}\)
 \\ \hline
   DFTB  & 1.01 & 0.19 & 0.15  & 0.52 & 0.39 & 0.12 & 0.36 & 0.45 & 0.12 & 0.34 & 0.42 & 0.13 & 0.19 & 0.46 & 0.15 \\
    LDA-LCAO-M & 0.97 & 0.08 & 0.53 & 0.56 & 0.14 & 0.18 & 0.39 & 0.18 & 0.20 & 0.33 & 0.16 & 0.22 & 0.20 & 0.26 & 0.14 \\
    LDA-LCAO-H & 1.00 & 0.04  & 0.79 & 0.57 & 0.13 & 0.26 & 0.41 & 0.18 & 0.26 & 0.34  & 0.16 & 0.27 & 0.21 & 0.22 & 0.18 \\
    LDA-LCAO-U & 0.99 & 0.04 & 0.81 & 0.56  & 0.13 & 0.26 & 0.40 & 0.18 & 0.27 & 0.33 & 0.16 & 0.26 & 0.21 & 0.22 & 0.18 \\
    LDA-PW & 0.64 & 0.39 & 0.31 & 0.17 & 0.54 & 0.08 & 0.10   & 0.50 & 0.11 & 0.08 & 0.44 & 0.09 & 0.01 & 0.41 & 0.02 \\
    RPBE-LCAO-M & 1.04 & 0.08 & 0.37 & 0.67 & 0.13  & 0.13  & 0.45 & 0.17 & 0.13 & 0.40 & 0.17 & 0.18 & 0.26 & 0.26 & 0.12 \\
    RPBE-LCAO-H & 1.06  & 0.04 & 0.53 & 0.67 & 0.12 & 0.18  & 0.47 & 0.17 & 0.17 & 0.41 & 0.15 & 0.22 & 0.26 & 0.22 & 0.18 \\
    RPBE-LCAO-U & 1.05 & 0.04  & 0.54 & 0.67 & 0.12 & 0.18 & 0.47 & 0.17 & 0.18 & 0.40 & 0.15 & 0.22  & 0.26 & 0.22 & 0.18 \\
    RPBE-PW & 0.75 & 0.33 & 0.23 & 0.28 & 0.52 & 0.07  & 0.16 & 0.49 & 0.08 & 0.14 & 0.43 & 0.08 & 0.03 & 0.49 & 0.02 \\
    PW91-LCAO-M & 1.01 & 0.08 & 0.28 & 0.63 & 0.13 & 0.11 & 0.43 & 0.18 & 0.11 & 0.38 & 0.17 & 0.15 & 0.24 & 0.26 & 0.11 \\
    PW91-LCAO-H & 1.04 & 0.04  & 0.58  & 0.63 & 0.12 & 0.20 & 0.45 & 0.17 & 0.19 & 0.39 & 0.16 & 0.23 & 0.25 & 0.23 & 0.17 \\
    PW91-LCAO-U & 1.03 & 0.04 & 0.59 & 0.63 & 0.12 & 0.20 & 0.45 & 0.17 & 0.19 & 0.38 & 0.16 & 0.22 & 0.25 & 0.22 & 0.17 \\
    PW91-PW & 0.73 & 0.33 & 0.23 & 0.25 & 0.53 & 0.07  & 0.14 & 0.50 & 0.08 & 0.12 & 0.44 & 0.08 & 0.02  & 0.48 & 0.02 \\
    PBE-LCAO-M & 1.02 & 0.08 & 0.31 & 0.64 & 0.13 & 0.12  & 0.44 & 0.17 & 0.12 & 0.38 & 0.17 & 0.16 & 0.24 & 0.26 & 0.12 \\
     PBE-LCAO-H & 1.04 & 0.04  & 0.59 & 0.65 & 0.12  & 0.20 & 0.46 & 0.17  & 0.19 & 0.39 & 0.15 & 0.23 & 0.25  & 0.22 & 0.17 \\
     PBE-LCAO-U & 1.04 & 0.04 & 0.60 & 0.64 & 0.12  & 0.20 & 0.45 & 0.17 & 0.19 & 0.39  & 0.15 & 0.23  & 0.25 & 0.22 & 0.18 \\
    PBE-PW & 0.73 & 0.33 & 0.23 & 0.25 & 0.53 & 0.07  & 0.14 & 0.50 & 0.08 & 0.12 & 0.44 & 0.08 & 0.02  & 0.49 & 0.02 \\ 
     B3LYP-LCAO-M & 0.62 & 0.10 & 0.01  & 0.41 & 0.14 & 0.00 & 0.29 & 0.18  & 0.00 & 0.27 & 0.15 & 0.00 & 0.19  & 0.22 & 0.01 \\
      B3LYP-LCAO-H & 0.81 & 0.03 & 0.02 & 0.55 & 0.10 & 0.01 & 0.41 & 0.15 & -0.01 & 0.37 & 0.13 & 0.00 & - & - & - \\
      B3LYP-LCAO-U & 0.78 & 0.03 & 0.02 & 0.55  & 0.10 & 0.01 & 0.41 & 0.15 & -0.01 & 0.37 & 0.14 & 0.00  & - & - & - \\
    B3LYP-PW & 0.55 & 0.26 & 0.02 & 0.21 & 0.43 & 0.01  & 0.13 & 0.40 & 0.02  & 0.12 & 0.35  & 0.02 & 0.04 & 0.39 & 0.01 \\
    PBE0-LCAO-M & 0.50 & 0.10 & 0.01 & 0.37 & 0.14 & 0.00 & 0.25 & 0.19  & 0.00 & 0.24 & 0.16 & 0.00  & - & - & - \\
  PBE0-LCAO-H & 0.71 & 0.03 & 0.02 & 0.51 & 0.10   & 0.01 & 0.38 & 0.15 & -0.01 & 0.34 & 0.13 & 0.00  & - & - & - \\
PBE0-LCAO-U & 0.70   & 0.03 & 0.02 & 0.50 & 0.10 & 0.01 & 0.37 & 0.15 & -0.01 & 0.34 & 0.13 & 0.00 & - & - & - \\
      PBE0-PW & 0.47 & 0.26 & 0.01  & 0.18 & 0.42  & 0.01 & 0.10 & 0.39  & 0.02 & 0.10 & 0.34 & 0.01 & 0.01 & 0.38  & 0.01 \\
  HSE03-LCAO-M & 0.92 & 0.07 & 0.03 & 0.58 & 0.12 & 0.03 & 0.39 & 0.16 & 0.02 & 0.35 & 0.15  & 0.04 & 0.23  & 0.25 & 0.07 \\
 HSE03-LCAO-H & 0.94 & 0.04 & 0.05 & 0.57 & 0.12 & 0.05 & 0.40 & 0.16 & 0.04 & 0.35 & 0.14 & 0.06  & - & - & - \\
 HSE03-LCAO-U & 0.93 & 0.04 & 0.05 & 0.57 & 0.12 & 0.05 & 0.40   & 0.16 & 0.04 & 0.35 & 0.14 & 0.06  & - & - & - \\
HSE03-PW & 0.67 & 0.30 & 0.03 & 0.22 & 0.49  & 0.01 & 0.13 & 0.45 & 0.02 & 0.11 & 0.40 & 0.02 & 0.02 & 0.46 & 0.01 \\
     HSE06-LCAO-M & 0.88 & 0.07 & 0.03 & 0.55 & 0.11 & 0.03 & 0.38 & 0.15 & 0.02 & 0.34 & 0.14 & 0.04 & 0.22 & 0.24 & 0.06 \\
     HSE06-LCAO-H & 0.90   & 0.03 & 0.04 & 0.55  & 0.11 & 0.04 & 0.39 & 0.15 & 0.04 & 0.34 & 0.13  & 0.05  & - & - & - \\
       HSE06-LCAO-U & 0.89 & 0.03 & 0.04 & 0.55 & 0.11 & 0.05 & 0.39 & 0.15 & 0.04  & 0.34 & 0.13  & 0.05  & - & - & - \\
    HSE06-PW & 0.63 & 0.29 & 0.02 & 0.21 & 0.47 & 0.01 & 0.12 & 0.43 & 0.02 & 0.11 & 0.38 & 0.02 & 0.02 & 0.45 & 0.01 \\
\end{tabular}
     \end{ruledtabular}

 \end{table*}
 
\end{document}